# Cooperative IoT Data Sharing with Heterogeneity of Participants Based on Electricity Retail

Bohong Wang, *Graduate Student Member, IEEE*, Qinglai Guo, *Senior Member, IEEE*, Tian Xia, *Member, IEEE*, Qiang Li, Di Liu, and Feng Zhao

*Abstract*—With the development of Internet of Things (IoT) and big data technology, the data value is increasingly explored in multiple practical scenarios, including electricity transactions. However, the isolation of IoT data among several entities makes it difficult to achieve optimal allocation of data resources and convert data resources into real economic value, thus it is necessary to introduce the IoT data sharing mode to drive data circulation. To enhance the accuracy and fairness of IoT data sharing, the heterogeneity of participants is sufficiently considered, and data valuation and profit allocation in IoT data sharing are improved based on the background of electricity retail. Data valuation is supposed to be relevant to attributes of IoT data buyers, thus risk preferences of electricity retailers are applied as characteristic attributes and data premium rates are proposed to modify data value rates. Profit allocation should measure the marginal contribution shares of electricity retailers and data brokers fairly, thus asymmetric Nash bargaining model is used to guarantee that they could receive reasonable profits based on their specific contribution to the coalition of IoT data sharing. Considering the heterogeneity of participants comprehensively, the proposed IoT data sharing fits for a large coalition of IoT data sharing with multiple electricity retailers and data brokers. Finally, to demonstrate the applications of IoT data sharing in smart grids, case studies are utilized to validate the results of data value for electricity retailers with different risk preferences and the efficiency of profit allocation using asymmetric Nash bargaining model.

## NOMENCLATURE

### Indices
| | |
|---|---|
| $i,j,k$ | Index of data brokers |
| $q,u$ | Index of electricity retailers |
| $l,m$ | Index of kernel functions |
| $t$ | Index of time periods |

### Parameters
| | |
|---|---|
| $X_{min}$ | Minimum value of real-time (RT) load demand (kWh) |
| $X_{max}$ | Maximum value of RT load demand (kWh) |
| $\sigma$ | Standard deviation of RT load demand |

This work was supported by the National Key Research and Development Program of China 2020YFB2104500. *(Corresponding author: Qinglai Guo)*

B. Wang, Q. Guo, T. Xia are with the Department of Electrical Engineering, Tsinghua University, Beijing, China. E-mail: wbh19@mails.tsinghua.edu.cn; guoqinglai@tsinghua.edu.cn; summersummer@mail.tsinghua.edu.cn.

Q. Li, D. Liu, F. Zhao are with State Grid Info & Telecom Group, Beijing, China. E-mail: {qiang-li, liudi, feng_zhao}@sgitg.sgcc.com.cn.

### Variables
| | |
|---|---|
| $\pi_t$ | Day-ahead (DA) market price in time period $t$ ($/kWh) |
| $p_t$ | Flat retail market price in time period $t$ ($/kWh) |
| $\lambda_t^+$ | Positive imbalance RT balancing market price in time period $t$ ($/kWh) |
| $\lambda_t^-$ | Negative imbalance RT balancing market price in time period $t$ ($/kWh) |
| $\Pi_t$ | Total profits of the electricity retailer in electricity transactions in time period $t$ ($) |
| $\tilde{X}_t$ | DA electricity bid of the electricity retailer in time period $t$ (kWh) |
| $X_t$ | RT electricity random variable of load demand of the electricity retailer in time period $t$ (kWh) |
| $x_t$ | RT electricity load demand of the electricity retailer in time period $t$ (kWh) |
| $\gamma_t^{*,a*,p*}$ | Optimal quantile of risk-neutral, averse, prone electricity retailer in time period $t$ (kWh) |
| $\tilde{X}_t^{*,a*,p*}$ | Optimal DA electricity bid of risk-neutral, averse, prone electricity retailer in time period $t$ (kWh) |
| $\eta^{a,p}$ | Degree of risk aversion, degree of risk proneness |
| $\varepsilon^{a,p}$ | Conservative quantile, aggressiveness quantile |
| $V_{\sigma,q}$ | Data value rate in electricity transactions for the $q$th electricity retailer ($/kWh) |
| $P_{\sigma,q}^{a,p}$ | Risk-averse, prone data premium rate in electricity transactions for the $q$th electricity retailer ($/kWh) |
| $R_{\sigma,q}$ | Marginal revenue from electricity transactions for the $q$th electricity retailer ($/kWh) |
| $\beta_i$ | Privacy sensitivity of end users related to the $i$th data broker ($/kWh) |
| $S_q$ | Total surplus of the $q$th electricity retailer and its data brokers ($) |
| $\sigma_{e,q}$ | Standard deviation of RT load demand for the $q$th electricity retailer |
| $\sigma_i$ | Standard deviation that the $i$th data broker can reduce (kWh) |
| $\Delta\sigma_{i,q}$ | Standard deviation that the $i$th data broker reduce for the $q$th electricity retailer (kWh) |
| $e_i$ | Cost-effectiveness index of the $i$th data broker (($/kWh)^{-0.5}) |
| $e_{si,q}$ | Cost-effectiveness criterion of the $i$th data broker for the $q$th electricity retailer (($/kWh)^{-0.5}) |
| $\tau_q^{er},\tau_i^{db}$ | Marginal profit shares of the $q$th electricity retailer and the $i$th data broker |
| $D_q^{er},D_i^{db}$ | Total revenues received of the $q$th electricity retailer and the $i$th data broker ($) |

| $U_C$ | Utility of the coalition from data sharing |
|---|---|
| $S_q^*, S_{q,\bar{i}}^*$ | Maximum total surplus of the $q$th electricity retailer and without the $i$th data broker (\$) |

*Functions*

| $f_{X_t}(x_t)$ | Probability density function (PDF) of RT load demand |
|---|---|
| $F_{X_t}(x_t)$ | Cumulative distribution function (CDF) of RT load demand |

## I. INTRODUCTION

AS Internet of Things (IoT) and big data technology advances, IoT data generated from numerous IoT devices have become one of the critical assets for market entities. Given that data may have profound value, data are considered among the five most important production factors in parallel with land, labor, capital, and technology [1]. As newly emerged production factors, data play an important role in several technology- and energy-intensive industries [2], to reduce energy consumption and enhance benefits. However, IoT data are generally stuck in the "data islands" of their producers and do not have mobility among market entities as expected, especially in power systems. Predictably, the lack of mobility makes it difficult to determine and realize the economic value of IoT data. For instance, the uncertainty of wind power [3] will bring incremental costs for dispatch to power systems and economic losses to renewable energy market participants, while improving forecasting accuracy would be an effective method. Wind turbine data [4], reference wind mast data, and numerical weather predictions [5] can improve wind energy forecasting. However, the entities who need IoT data and the entities who produce or possess IoT data are usually not the same. Wind power plant operators who need more accurate wind energy forecasting hope to obtain wind turbine data from wind turbine manufacturers, reference wind mast data, and numerical weather predictions from the government meteorological department. If wind turbine manufacturers or the government meteorological department are unwilling to provide relevant data to wind power plant operators, wind power plant operators cannot obtain extra benefits. Hence, in a sense, the mobility of IoT data is more important than the value of IoT data.

To increase the mobility of IoT data and construct a framework for IoT data interchange, the works of current literature can generally be divided into two major aspects: data privacy protection and data sharing [6]. On the one hand, data privacy needs to be well protected, which allows market entities not to worry about inappropriate privacy exposure; on the other hand, data sharing needs to be reasonable and fair so that the monetary incentives that market entities receive can make them willing to actively participate in data sharing. Data privacy protection is the more practiced and studied of these two major aspects. The omission of data privacy protection poses a serious threat to the application of data sharing, and multiple countries worldwide have paid attention to data privacy protection in their policies [7], [8]. Comparatively, there are few studies on data sharing design, and their theoretical framework is not sufficiently systematic, especially in the field of power systems. Data sharing should be composed of two stages: the first stage is to identify and predict the total revenues and costs related to data sharing; the second stage is to take the difference in revenues and costs as the surplus and to allocate it to market entities that contribute to data sharing.

The first stage can be defined as "data valuation". Xu *et al.* [9] focused on the mobile health applications, used the differential entropy of model parameters' distributions to measure the data's contribution to the enhancement of prediction accuracy, and proposed an online data valuation metric under the Bayesian perspective. Yu *et al.* [10] proposed a data valuation paradigm based on Shannon entropy [11] and the non-noise ratio, constructed the functional relationships between prediction accuracy and the two parameters, and validated the paradigm using the conventional unit commitment (UC) model and real photovoltaic forecasting data. Wang *et al.* [12] proposed a data valuation paradigm based on the two-settlement mechanism in electricity transactions and derived formulas of data value rates which indicate the profit enhancement that corresponds to unit uncertainty reduction brought by data. Existing studies emphasized on the data valuation based on the inherent attributes of data themselves. However, unlike ordinary commodities, data are usually with the value that is affected by the application entities and scenarios of data.

The second stage can be defined as "profit allocation". Referring to the data marketplace with a robust real-time matching mechanism in [13], Gonçalves *et al.* [14] used Myerson's payment function [15] and Shapley fairness, extended the model for a sliding window environment, and no longer distinguished between data buyers and sellers. Han *et al.* [16] demonstrated that users' schedulable load data can improve the load forecasting of the energy retailer's profits and used the Shapley value and the nucleolus in the model of the cooperative game between the energy retailer and users. Wang *et al.* [17] proposed a novel profit allocation mechanism to distribute the total consumer surplus obtained by the electricity retailer and data suppliers in data sharing, where the closed-form formulas lead to shorter computation time and higher profitability for the electricity retailer than Shapley value method. Existing studies always separate the profit settlement of the data buyer-side and profit allocation of the data seller-side. However, as the data sharing involves data buyers and sellers, their heterogeneous contributions should be simultaneously considered.

Focusing on the flaws mentioned above, some critical contributions are proposed in this paper based on our previous work [12], [17]. Electricity transactions of electricity retailers are still applied as a typical scenario for our analysis, where electricity retailers and data brokers serve as buyers and sellers of data services in a coalition of IoT data sharing to share the profits generated from IoT data of end users. Moreover, operational processes of IoT data sharing are roughly similar to that in [17], necessary means to protect the security of IoT data and power systems and ex-post supervisions of results of data valuation and profit allocation may also be needed. For data security, distributed ledger technologies (DLTs) and blockchains [18] can be used to record the details of IoT data sharing history and guarantee the transparency for IoT data, and digital wa-termarking [19] on IoT data may protect end users' copyright and deter data brokers from data privacy disclosure. Besides, a third-party regulatory platform conduct security checks of power systems after IoT data sharing and electricity market clearing for security of IoT data and power systems, and it can also regularly check the results and impose corresponding penalties on behavior and market participants that violate regulations for ex-post supervisions. To fill the gap in data valuation, risk preference, which is usually discussed in decision-making problems, is selected as the characteristic attribute of data buyers [20], [21] to depict their differences. The risk preference can be obtained from the bidding decision processes of market participants through inverse reinforcement learning (IRL) [22], [23]. To the best of our knowledge, few studies have discussed the differences of data value for different data buyers. Besides, to address the challenge that profit allocation merely for data sellers are not suitable enough for IoT data sharing, the basis for profit allocation should be reasonably determined according to the heterogeneous contributions of data buyers and sellers to the coalition, where data buyers create the demand of IoT data sharing and provide revenue sources, while data sellers meet the demand and bring incremental economic value.

The main contributions of this paper are as follows:

a) A data valuation model simultaneously incorporating different uncertainty-reducing effects of data and different characteristic attributes of electricity retailers is constructed, where risk preferences as characteristic attributes will affect the decision-making of electricity retailers and bring the data premium beyond data value.

b) A profit allocation method using asymmetric Nash bargaining model is employed to consider heterogeneous contributions of electricity retailers and data brokers in the coalition of IoT data sharing, and electricity retailers and data brokers will receive their profits according to pre-defined marginal profit shares.

The remainder of this paper is organized as follows. Based on our previous work, electricity transactions as background are briefly introduced in Section II, where optimization models for electricity retailers with different risk preferences are solved and data valuation is improved. According to the solution of consumer surplus maximization, an asymmetric Nash bargaining model for profit allocation between electricity retailers and data brokers is described in Section III. In Section IV, case studies are used to illustrate results of data valuation and profit allocation. Finally, Section V concludes this paper and briefly introduces three potential research directions for future work.

## II. DATA VALUATION WITH RISK PREFERENCES OF ELECTRICITY RETAILERS

### A. Background

Electricity retailers are the main market entities on the consumption side of power systems and serve as agents for many medium- and small-sized end users to purchase electricity through electricity transactions. Although electricity retailers bear important responsibilities in electricity consumption, they also directly face risks from electricity wholesale markets and end users. On the one hand, there are numerous electricity retailers, and each of them has too little market power to have a significant impact on market clearing prices; thus, they have to bear the fluctuation risks of electricity market prices. On the other hand, it seems that the rights for end users to choose electricity retailers are substantially greater than the rights for electricity retailers to choose end users [24], [25]. As service providers, electricity retailers need to improve their service quality as much as possible to obtain the recognition of end users rather than requiring end users to achieve specific electricity consumption goals. Hence, electricity retailers must bear the uncertainty of end users' load demand.

Comparing the risks from electricity wholesale markets and end users, the uncertainty of end users' load demand is a greater threat to electricity retailers' profits because electricity wholesale markets have more explicit operating patterns and their prices can be accurately forecasted [26], [27]. To reduce the uncertainty in electric load forecasting, electricity retailers are willing to obtain some end users' IoT data (e.g. smart meter data) to derive useful information. Recent studies on load forecasting emphasize the significance of data and further improve their accuracy by better dealing with uncertainty [28], [29], which proves the feasibility of the scenario in this paper. Note that IoT data usually have low value density; that is, large amounts of IoT data may yield trivial value; thus, it is somewhat uneconomical for electricity retailers to collect IoT data themselves. Fortunately, data brokers have become important market entities that link the data sources and the data demand of electricity retailers [30]. Data brokers reap revenues from the processes of collecting raw IoT data from end users, processing them, providing data products or services, and assisting electricity retailers in increasing their profits.

Given certain characteristics of electricity retail and that data have replicability and non-excludability, the data transactions between electricity retailers and data brokers differ from common transactions of commodities in terms of the following:

a) data transactions between electricity retailers and data brokers are probably not one-time but continuous because end users' IoT data are time-varying;

b) the transaction relationships between one electricity retailer and one data broker are highly flexible because end users can choose different electricity retailers in different time intervals according to electricity retailers' pricing schemes and other influencing factors to consume electricity at a more favorable price;

c) the transaction relationships between a group of electricity retailers and a group of data brokers are relatively fixed because electricity retailers are limited by geographical regions and data brokers would also expect that it is expensive to replace data sources.

Considering the characteristics listed above, an IoT data sharing coalition better depicts the relationship between electricity retailers and data brokers. The structure of the IoT data sharing coalition is shown in Fig. 1. In the coalition, data brokers collect optimal amounts of IoT data from end users and pay corresponding data costs, while electricity retailers bid optimal loads to electricity market operators and receive data revenues. The IoT data collected by data brokers translate into better decisions being made by electricity retailers through the improvement of load forecasting, and the profits will finally flow to those electricity retailers and data brokers who contribute to data sharing.

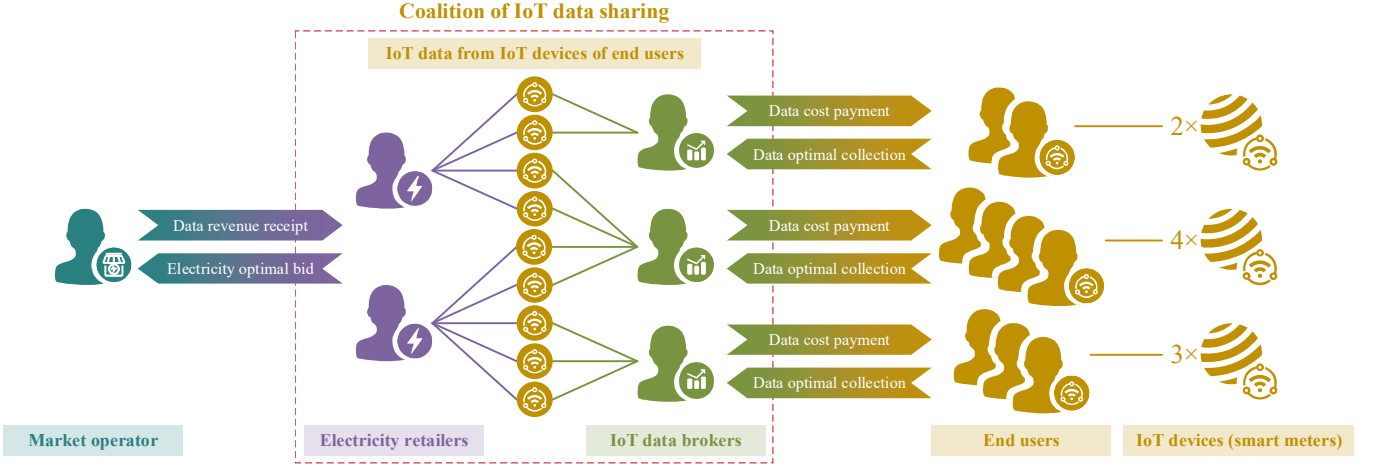

Fig. 1. The coalition of data sharing comprises of electricity retailers who participate in electricity transactions and receive data revenues and data brokers who collect data and pay data costs to end users.

### B. Optimization Models Based on Electricity Transactions

To achieve data valuation, it is necessary to model electricity transactions and determine the revenue sources. When participating in electricity wholesale markets, electricity retailers should bid in the two-settlement market system and will suffer considerable imbalance costs caused by inaccurate load forecasting. Under the two-settlement electricity market, the profits of the electricity retailer in time interval $t$ can be found in [12]. Classifying the terms with $\tilde{X}_t$ and the terms with $x_t$ to highlight the possible electricity imbalance and facilitate the subsequent derivation, the profits of the electricity retailer in time interval $t$ can be expressed as

$$\Pi_t = \begin{cases} \left(p_t - \lambda_t^+\right)x_t + \left(\lambda_t^+ - \pi_t\right)\tilde{X}_t, X_{\min} \le x_t < \tilde{X}_t \\ \left(p_t - \lambda_t^-\right)x_t + \left(\lambda_t^- - \pi_t\right)\tilde{X}_t, \tilde{X}_t \le x_t \le X_{\max} \end{cases}. \quad (1)$$

According to the analysis in Section II-A, electricity retailers will be regarded as price takers; thus, they can only enhance their profits by making their day-ahead electricity bid as close to the real-time electricity demand as possible instead of affecting electricity prices. In addition, the power flow constraints in power systems will not be included because the nonlinear optimal power flow model will obstruct data value discovery and distort the results of data valuation. To ensure the credibility of the data value derived, it can be assumed that power flows have not changed significantly in the decision-making of electricity retailers. Without regard to the power flow constraints, the optimization model for electricity retailers will be a single-objective model with an upper and lower bound constraint for their day-ahead electricity bid.

The optimization model for risk-neutral electricity retailers is expressed as [12]

$$\max_{\tilde{X}_t} \mathbb{E}\Pi_t\left(\pi_t, \lambda_t^+, \lambda_t^-\right)$$
$$= \int_{X_{\min}}^{X_{\max}} \Pi_t \cdot f_{X_t}(x_t)\mathrm{d}x_t$$
$$= \int_{X_{\min}}^{\tilde{X}_t} \left[\left(p_t - \lambda_t^+\right)x_t + \left(\lambda_t^+ - \pi_t\right)\tilde{X}_t\right]f_{X_t}(x_t)\mathrm{d}x_t$$
$$+ \int_{\tilde{X}_t}^{X_{\max}} \left[\left(p_t - \lambda_t^-\right)x_t + \left(\lambda_t^- - \pi_t\right)\tilde{X}_t\right]f_{X_t}(x_t)\mathrm{d}x_t$$

$$s.t. \quad X_{\min} \le \tilde{X}_t \le X_{\max}. \quad (3)$$

The first-order derivative of the objective function in Eq. (2) is

$$\frac{\mathrm{d}\mathbb{E}\Pi_t\left(\pi_t, \lambda_t^+, \lambda_t^-\right)}{\mathrm{d}\tilde{X}_t} = \left(\lambda_t^- - \pi_t\right) - \left(\lambda_t^- - \lambda_t^+\right)F_{X_t}(\tilde{X}_t). \quad (4)$$

Using the first-order optimality condition, the optimal quantile and the optimal day-ahead electricity bid can be obtained when the first-order derivative equals zero, where the numerical order of three market prices conforms to the supply-demand relationships of electricity markets in which the electricity retailer should sell excess electricity at lower prices or buy deficit electricity at higher prices in the real-time balancing market, namely, $\lambda_t^+ < \pi_t < \lambda_t^-$.

Since risk preferences are included in the discussion of the decision-making of electricity retailers, their optimization objectives will be differentiated. The objective function of risk-neutral electricity retailers is the expected profit, while the objective functions of risk-averse and risk-prone electricity retailers need to include risk adjustment terms in addition to the expected profit. Conditional value-at-risk (CVaR) is a commonly used consistent measure of risk [31], which originated in the field of investment and has also been extensively used in the field of electricity transactions [32]. In this paper, the risk preferences of electricity retailers are revealed by the CVaR terms in their objective functions. Note that CVaR is the mean excess loss or mean shortfall to measure left-side tail risks; thus, theoretically, it is only suitable for risk-averse electricity retailers who are more concerned about the lowest possible profits. Considering that the imbalance costs come from two sides of the electricity imbalance, CVaR is otherwise modified to measure the right-side tail risks to satisfy the requirement of risk-prone electricity retailers. Therefore, objective functions for risk-averse and risk-prone electricity retailers can be constructed as follows, and the constraints are also Eq. (3). To better compare optimization models for electricity retailers with different risk preferences, a graphical illustration of three optimization models under the assumption that the load forecasts are distributed uniformly is shown in Fig. 2.

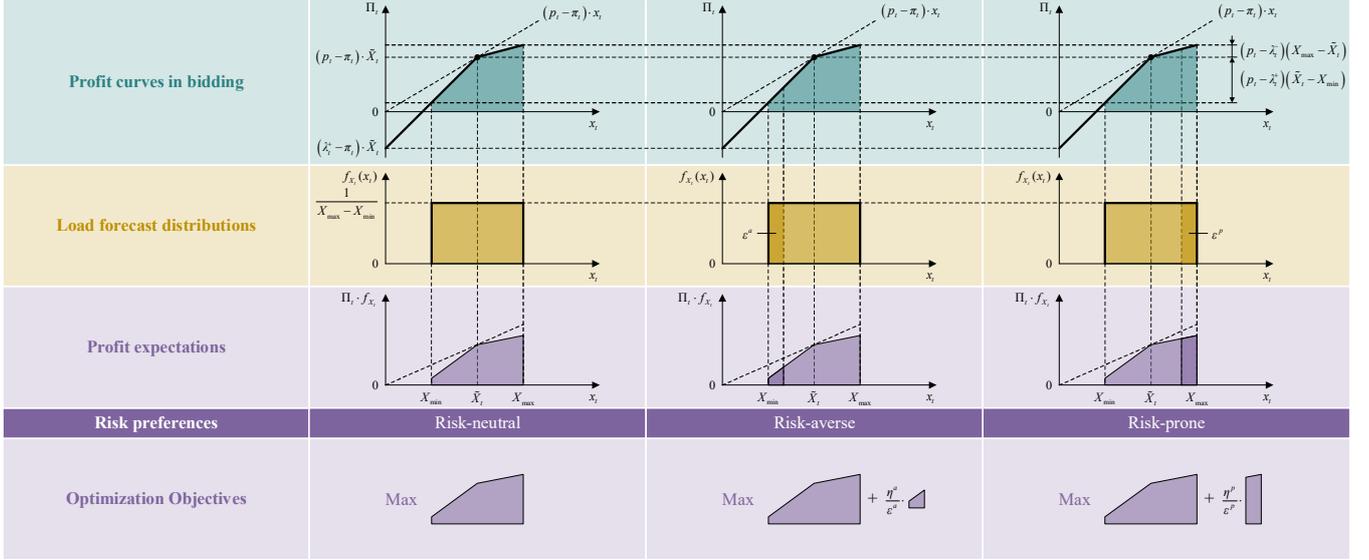

Fig. 2. Comparison of optimization models for risk-neutral, risk-averse, and risk-prone electricity retailers.

$$
\begin{aligned}
\max_{\tilde{X}_t} \ & \mathbb{E}^a \Pi_t \left( \pi_t, \lambda_t^+, \lambda_t^- \right) \\
&= \mathbb{E} \Pi_t \left( \pi_t, \lambda_t^+, \lambda_t^- \right) + \eta^a \mathrm{CVaR}_{\varepsilon^a} \\
&= \int_{X_{\min}}^{X_{\max}} \Pi_t \cdot f_{X_t}(x_t) \mathrm{d}x_t + \frac{\eta^a}{\varepsilon^a} \int_{X_{\min}}^{F_{X_t}^{-1}(\varepsilon^a)} \Pi_t \cdot f_{X_t}(x_t) \mathrm{d}x_t \\
&= \int_{X_{\min}}^{\tilde{X}_t} \left[ \left( p_t - \lambda_t^+ \right) x_t + \left( \lambda_t^+ - \pi_t \right) \tilde{X}_t \right] f_{X_t}(x_t) \mathrm{d}x_t \\
&\quad + \int_{\tilde{X}_t}^{X_{\max}} \left[ \left( p_t - \lambda_t^- \right) x_t + \left( \lambda_t^- - \pi_t \right) \tilde{X}_t \right] f_{X_t}(x_t) \mathrm{d}x_t \\
&\quad + \frac{\eta^a}{\varepsilon^a} \int_{X_{\min}}^{F_{X_t}^{-1}(\varepsilon^a)} \Pi_t \cdot f_{X_t}(x_t) \mathrm{d}x_t,
\end{aligned}
\tag{5}
$$

$$
\begin{aligned}
\max_{\tilde{X}_t} \ & \mathbb{E}^p \Pi_t \left( \pi_t, \lambda_t^+, \lambda_t^- \right) \\
&= \mathbb{E} \Pi_t \left( \pi_t, \lambda_t^+, \lambda_t^- \right) + \eta^p \mathrm{CVaR}_{\varepsilon^p}^+ \\
&= \int_{X_{\min}}^{X_{\max}} \Pi_t \cdot f_{X_t}(x_t) \mathrm{d}x_t + \frac{\eta^p}{\varepsilon^p} \int_{F_{X_t}^{-1}(1-\varepsilon^p)}^{X_{\max}} \Pi_t \cdot f_{X_t}(x_t) \mathrm{d}x_t \\
&= \int_{X_{\min}}^{\tilde{X}_t} \left[ \left( p_t - \lambda_t^+ \right) x_t + \left( \lambda_t^+ - \pi_t \right) \tilde{X}_t \right] f_{X_t}(x_t) \mathrm{d}x_t \\
&\quad + \int_{\tilde{X}_t}^{X_{\max}} \left[ \left( p_t - \lambda_t^- \right) x_t + \left( \lambda_t^- - \pi_t \right) \tilde{X}_t \right] f_{X_t}(x_t) \mathrm{d}x_t \\
&\quad + \frac{\eta^p}{\varepsilon^p} \int_{F_{X_t}^{-1}(1-\varepsilon^p)}^{X_{\max}} \Pi_t \cdot f_{X_t}(x_t) \mathrm{d}x_t.
\end{aligned}
\tag{6}
$$

The first-order derivatives of the objective functions in Eqs. (5) and (6) can also be obtained. Taking the objective function of risk-averse electricity retailers as an example, the first-order derivative of the objective function com-prises the first-order derivative of the expected profit, which is the same as Eq. (4), and the first-order derivative of the CVaR term multiplied by its coefficient $\eta^a$ as follows:

$$
\frac{\mathrm{d}\mathbb{E}^a \Pi_t \left( \pi_t, \lambda_t^+, \lambda_t^- \right)}{\mathrm{d}\tilde{X}_t} = \frac{\mathrm{d}\mathbb{E} \Pi_t \left( \pi_t, \lambda_t^+, \lambda_t^- \right)}{\mathrm{d}\tilde{X}_t} + \eta^a \frac{\mathrm{d}\mathrm{CVaR}_{\varepsilon^a}}{\mathrm{d}\tilde{X}_t}. \tag{7}
$$

It is noted that the CVaR term in Eq. (7) is merely related to the part of the load demand distribution that is lower than the $\varepsilon^a$-quantile, while the quantile of the day-ahead electricity bid cannot be predetermined; thus, whether the quantile of the day-ahead electricity bid is larger than $\varepsilon^a$ is uncertain without the solution of the optimization model. The first-order derivative of the CVaR term needs a detailed discussion since the profit curve in Eq. (1) is a piecewise function whose piecewise point is the quantile of the day-ahead electricity bid as follows:

$$
\begin{aligned}
\frac{\mathrm{d}\mathrm{CVaR}_{\varepsilon^a}}{\mathrm{d}\tilde{X}_t} &= \frac{1}{\varepsilon^a} \frac{\mathrm{d}\int_{X_{\min}}^{F_{X_t}^{-1}(\varepsilon^a)} \Pi_t \cdot f_{X_t}(x_t) \mathrm{d}x_t}{\mathrm{d}\tilde{X}_t} \\
&= \begin{cases} \lambda_t^+ - \pi_t & , \varepsilon^a \leq F_{X_t}(\tilde{X}_t) \\ \lambda_t^- - \pi_t - \dfrac{\lambda_t^- - \lambda_t^+}{\varepsilon^a} F_{X_t}(\tilde{X}_t) & , \varepsilon^a > F_{X_t}(\tilde{X}_t) \end{cases}.
\end{aligned}
\tag{8}
$$

Combining Eq. (4) and Eq. (8), the derivative of the objective functions in Eq. (5) can be obtained. Following the similar process, the first-order derivatives of the CVaR term and the objective function in Eq. (6) can be derived as

$$
\begin{aligned}
\frac{\mathrm{d}\mathrm{CVaR}_{\varepsilon^p}^+}{\mathrm{d}\tilde{X}_t} &= \frac{1}{\varepsilon^p} \frac{\mathrm{d}\int_{F_{X_t}^{-1}(1-\varepsilon^p)}^{X_{\max}} \Pi_t \cdot f_{X_t}(x_t) \mathrm{d}x_t}{\mathrm{d}\tilde{X}_t} \\
&= \begin{cases} \lambda_t^- - \pi_t + \dfrac{\lambda_t^- - \lambda_t^+}{\varepsilon^p} \left( 1 - F_{X_t}(\tilde{X}_t) \right), 1 - \varepsilon^p \leq F_{X_t}(\tilde{X}_t) \\ \lambda_t^- - \pi_t & , 1 - \varepsilon^p > F_{X_t}(\tilde{X}_t) \end{cases}.
\end{aligned}
\tag{9}
$$

As mentioned earlier, $\gamma_t^\star$ is limited in the interval (0,1) because of the supply-demand relationship in electricity markets.

However, for risk-averse electricity retailers, their optimal quantile for electricity bid will be an offset compared to that of risk-neutral electricity retailers due to their risk preferences. Moreover, $\gamma_t^{a*}$ must be located in the interval $(0,1)$ to make the formulas of the optimal quantile and optimal day-ahead electricity bid meaningful; otherwise, the optimal day-ahead electricity bid will be stuck on the upper or lower bound of the constraint. Hence, the optimal quantile of risk-averse electricity retailers can be expressed as a piecewise function:

$$\gamma_t^{a*} = \begin{cases} 0 & , 0 < \gamma_t^* < \dfrac{\eta^a}{1+\eta^a} \\[2mm] \left(1+\eta^a\right)\gamma_t^* - \eta^a & , \dfrac{\eta^a}{1+\eta^a} \le \gamma_t^* < \dfrac{\eta^a + \varepsilon^a}{1+\eta^a} \\[2mm] \dfrac{\varepsilon^a\left(1+\eta^a\right)}{\varepsilon^a + \eta^a}\gamma_t^* & , \dfrac{\eta^a + \varepsilon^a}{1+\eta^a} \le \gamma_t^* < 1 \end{cases} \quad (10)$$

Similarly, the optimal quantile of risk-prone electricity retailers can also be expressed as a piecewise function:

$$\gamma_t^{p*} = \begin{cases} \left(1+\eta^p\right)\gamma_t^* & , 0 \le \gamma_t^* < \dfrac{1-\varepsilon^p}{1+\eta^p} \\[2mm] \dfrac{\varepsilon^p\left(1+\eta^p\right)}{\varepsilon^p + \eta^p}\gamma_t^* - \dfrac{\eta^p\left(1-\varepsilon^p\right)}{\varepsilon^p + \eta^p} & , \dfrac{1-\varepsilon^p}{1+\eta^p} < \gamma_t^* \le 1 \end{cases} \quad (11)$$

The relationships between optimal quantiles for risk-averse and risk-prone electricity retailers with respect to the optimal quantile for risk-neutral electricity retailers are depicted in Fig. 3, where it can be found that risk preferences significantly affect the decision-making of electricity retailers. The optimal quantiles of risk-averse electricity retailers are consistently lower than those of risk-neutral electricity retailers, which indicates that their decision-making is more conservative; the optimal quantiles of risk-prone electricity retailers are consistently higher than those of risk-neutral electricity retailers, which indicates that their decision-making is more aggressive. Note that the risk preference of one electricity retailer could be quite different in different time intervals and different perceptions of electricity markets; thus, the parameters should be adjusted as required to more accurately depict the risk preferences of electricity retailers in their decision-making for the next time interval. Finally, the optimal day-ahead electricity bids for electricity retailers can be naturally obtained through the inverse function of the cumulative distribution function (CDF) of load demand.

### C. Data Valuation with Risk Preferences

Based on the solutions of optimization models, the optimal expected profits of electricity retailers will be derived. Referring to [12], the optimal expected profits of risk-neutral electricity retailers can be expressed in an affine-form formula as follows:

$$\mathbb{E}\Pi_t^* = \left(p_t - \pi_t\right)\mathbb{E}X_t - V_\sigma \cdot \sigma. \quad (12)$$

where the coefficient of $\sigma$ is defined as the data value rate and proved to be constant given the market price data in [12].

For risk-averse and risk-prone electricity retailers, the optimal expected profits are obtained by substituting their optimal day-ahead electricity bids into the formula for the expected profits instead of their objective functions with CVaR terms:

$$\mathbb{E}\Pi_t^{a*} = \left(p_t - \pi_t\right)\mathbb{E}X_t - \left(V_\sigma + P_\sigma^a\right)\cdot \sigma, \quad (13)$$

$$\mathbb{E}\Pi_t^{p*} = \left(p_t - \pi_t\right)\mathbb{E}X_t - \left(V_\sigma + P_\sigma^p\right)\cdot \sigma. \quad (14)$$

It is shown that their coefficients of $\sigma$ can be regarded as the sum of the data value rate and another term. The additional terms for risk-averse and risk-prone electricity retailers are interpreted as risk-averse and risk-prone data premium rates in this paper, whose formulas are symmetric and as follows:

$$P_\sigma^a = \left(\lambda_t^- - \lambda_t^+\right)\frac{\int_{\tilde{X}_t^{a*}}^{\tilde{X}_t^*} x_t f_{X_t}(x_t)\mathrm{d}x_t - \left(\gamma_t^* - \gamma_t^{a*}\right)\tilde{X}_t^{a*}}{\sigma}$$

$$= \left(\lambda_t^- - \lambda_t^+\right)\frac{\int_{\tilde{X}_t^{a*}}^{\tilde{X}_t^*} \left(\gamma_t^* - F_{X_t}(x_t)\right)\mathrm{d}x_t}{\sigma}, \quad (15)$$

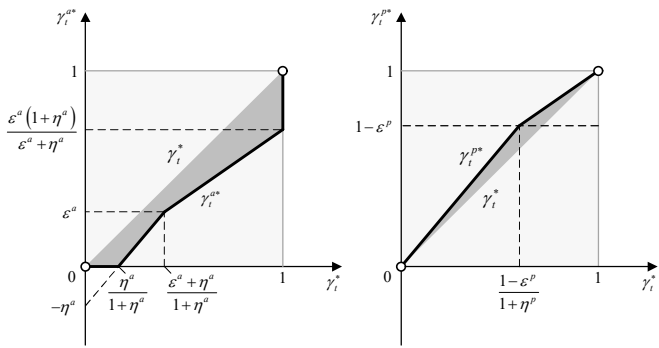

Fig. 3. Relationships of optimal quantiles for risk-averse and risk-prone electricity retailers with respect to the optimal quantile for risk-neutral electricity retailers.

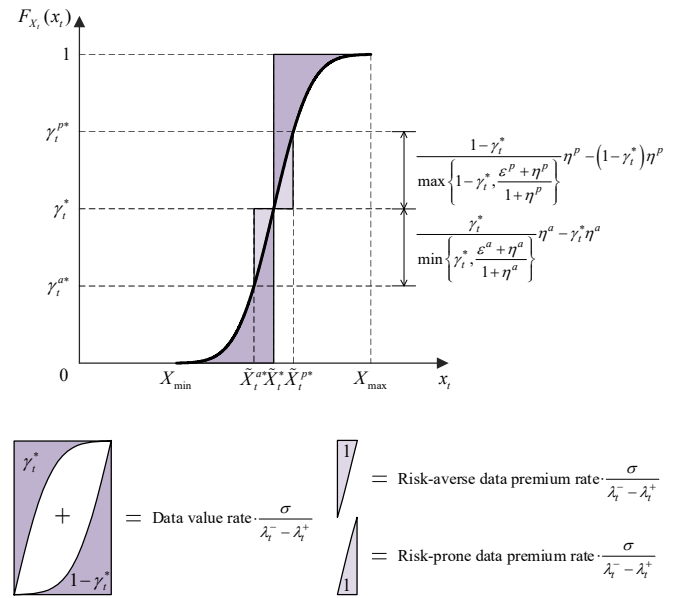

Fig. 4. Graphical illustration of data value rate, risk-averse data premium rate, and risk-prone data premium rate.

$$P_\sigma^p = \left(\lambda_t^- - \lambda_t^+\right) \frac{\int_{\tilde{X}_t^-}^{\tilde{X}_t^{p*}} x_t f_{X_t}(x_t)\mathrm{d}x_t - \left(\gamma_t^{p*} - \gamma_t^*\right)\tilde{X}_t^{p*}}{\sigma}$$

$$= \left(\lambda_t^- - \lambda_t^+\right)\frac{\int_{\tilde{X}_t^-}^{\tilde{X}_t^{p*}}\left(F_{X_t}(x_t) - \gamma_t^*\right)\mathrm{d}x_t}{\sigma}. \quad (16)$$

The data premiums indicate that data with the same effect on uncertainty reduction are more valuable for electricity retailers with specific risk preferences than for risk-neutral electricity retailers. In addition, the result that the risk-averse data premium rate and risk-prone data premium rate are both positive is somewhat counterintuitive. It can be understood that risk-averse and risk-prone electricity retailers both hope to benefit from their specific handling of risks, but better handling of risks requires more data to more accurately predict the risk level. Therefore, regardless of risk preferences, data premium rates for electricity retailers will always be positive.

To more clearly display the relationships among the data value rate, risk-averse data premium rate, and risk-prone data premium rate, their graphical illustrations are shown in Fig. 4 under the assumption that the load forecast is distributed normally. Besides, the complete calculation processes of data valuation can be summarized.

---

**Calculation process 1:** Data valuation with risk preferences (for electricity retailers)

1. **Inputs:** Given the day-ahead end users' electricity load forecast $\tilde{X}_t$ and day-ahead electricity market price $\pi_t$, electricity retailers set their own retail price $p_t$.
2. **Assumptions:**
   a) The positive and negative penalty rates $\lambda_t^+$ and $\lambda_t^-$ for electricity imbalance are proportional to the day-ahead electricity market price $\pi_t$;
   b) Electricity retailers can predict the penalty rates relatively accurately compared with their forecast for real-time electricity load demand.
3. Choose parametric estimation or nonparametric estimation to depict the forecast errors of real-time electricity load demand.
4. Optimize the estimated parameters (*e.g.* $\mu$, $\sigma$) of the distribution of real-time electricity load demand in through minimizing the estimation error.
5. Obtain the estimated distribution based on the estimated parameters.
6. Calculate the data value rates according to the derived formulas correspond to different estimated distributions.
7. Calculate the data premium rates according to the derived formulas with respect to the risk preferences of electricity retailers.
8. **Outputs:** Data value rate $V_{\sigma,q}$, risk-averse data premium rate $P_{\sigma,q}^a$ or risk-prone data premium rate $P_{\sigma,q}^q$.

---

## III. PROFIT ALLOCATION USING ASYMMETRIC NASH BARGAINING

### A. Data Profits

The marginal data cost for each data broker is considered to increase exponentially in the reduction in uncertainty $\Delta\sigma_i$, which indicates that to reduce unit uncertainty, the payment for collecting data will increase exponentially as the uncertainty level approaches zero. The payment for collecting data is proportional to the amount of data collected, and the exponential relationship can be interpreted as follows: the information contained in a large amount of data has many duplicates; thus, exponentially more data are needed to mine incremental information to further reduce uncertainty. The two main parameters for the data broker are $\sigma_i$ and $\beta_i$. $\sigma_i$ is the maximum reduction in uncertainty that the data broker can achieve, but it should be noted that the uncertainty can never be completely eliminated because, on the one hand, there are always some private data of end users that cannot be obtained, and on the other hand, load forecasts can never be absolutely accurate. $\beta_i$ is the privacy sensitivity of end users that the data broker connects with and will affect the steepness of the exponential curve because when the privacy sensitivity of end users is higher, the same data are more difficult for the data broker to collect; thus, the exponential curve will be steeper, and the breakeven point for electricity retailers and data brokers will come sooner in the process of data collection.

In addition, the formula for the marginal data revenue can be obtained based on the data valuation in Section II-C:

$$R_{\sigma,q} = \begin{cases} V_{\sigma,q} & , \forall q \in \mathcal{R}^n \\ V_{\sigma,q} + P_{\sigma,q}^a, \forall q \in \mathcal{R}^a \\ V_{\sigma,q} + P_{\sigma,q}^p, \forall q \in \mathcal{R}^p \end{cases}. \quad (17)$$

For the coalition, the marginal data revenue is a constant determined by market prices and the risk preferences of electricity retailers regardless of uncertainty, while the marginal data cost is a variable related to the degree of uncertainty reduction. Given the formulas for the marginal data revenue and the marginal data cost, data profits within the data sharing coalition can be analyzed. From the perspective of market analysis, the total surplus of the coalition is the product of the "volume" and the "unit price", where the "volume" is the degree of uncertainty reduction in mathematical expressions, and the "unit price" is the difference between the marginal data revenue and marginal data cost. If the coalition as the data demander and end users as the data suppliers are in a perfectly competitive data market, the market equilibrium will lead to the marginal data revenue being equal to the marginal data cost; thus, the surplus for the coalition will be zero since the marginal data revenue is a constant. Even if it is not a perfectly competitive data market but a data market where end users have a certain market power, the surplus of the coalition will be greatly depressed. Therefore, it is assumed that end users are passive participants who fit the description of the formula for the marginal data cost. The rationality of this assumption is that not all end users care about the market value of their private data and using the private data of some end users can reduce a large part of uncertainty; thus, the fact that the supply of privacy data is far less than their demand makes it difficult for a single end user to obtain excess returns by selling his or her private data. Under this assumption, the electricity retailers in the coalition will make the data collection decision by maximizing the total surplus of their data brokers and themselves, and the optimization model for the $q$th electricity retailer is expressed as

$$\max_{\Delta\sigma_{i,q}} S_q = R_{\sigma,q}\left(\sigma_{e,q} - \sqrt{\sum_{i\in\mathcal{B}}\left(\sigma_i - \Delta\sigma_{i,q}\right)^2}\right) \\ - \sum_{i\in\mathcal{B}}\beta_i\ln\frac{\sigma_i}{\sigma_i - \Delta\sigma_{i,q}} \tag{18}$$

$$s.t. \quad 0 \le \Delta\sigma_{i,q} < \sigma_i, \forall i \in \mathcal{B}. \tag{19}$$

According to the analysis above, the key to maximizing the total surplus is to select the most appropriate level of uncertainty reduction given that the marginal data revenue is fixed. Hence, the cost-effectiveness of data brokers is important because data brokers who are highly cost-effective will be included in the coalition, where high cost-effectiveness implies that data brokers can achieve a relatively high level of uncertainty reduction under a low privacy sensitivity of end users. To better express the solution of the optimization model, the measure of data brokers' cost-effectiveness is defined as follows [17]:

$$e_i = \frac{\sigma_i}{\sqrt{\beta_i}}, \forall i \in \mathcal{B}. \tag{20}$$

Following the arrangement in [17], the cost-effectiveness index $e_i$ of data brokers is in a non-ascending order with respect to their serial numbers, which does not affect the generality of the optimal solution:

$$e_i \ge e_j, \forall i,j \in \mathcal{B}, i < j. \tag{21}$$

To solve the optimization model, two key propositions are proposed and proven as follows:

*Proposition 1:* If the $k$th data broker is an active data broker, i.e., $\Delta\sigma_{i,q} > 0$ in the optimal solution, any $i$th data broker with a smaller serial number ($i < k$) will be an active data broker; if it is an inactive data broker, i.e., $\Delta\sigma_{i,q} = 0$ in the optimal solution, any $i$th data broker with a larger serial number ($i > k$) will be an inactive data broker, namely,

$$\Delta\sigma_{k,q} > 0 \Rightarrow \Delta\sigma_{i,q} > 0, \forall i,k \in \mathcal{B}, i < k, \forall q \in \mathcal{R}, \tag{22}$$

$$\Delta\sigma_{k,q} = 0 \Rightarrow \Delta\sigma_{i,q} = 0, \forall i,k \in \mathcal{B}, i > k, \forall q \in \mathcal{R}. \tag{23}$$

*Proof:* Refer to Appendix A for a detailed proof.

*Proposition 2:* The $i$th data broker is classified as an active data broker for the $q$th electricity retailer if and only if its cost-effectiveness index satisfies the following criterion:

$$e_i = \frac{1}{\sqrt{2}R_{\sigma,q}}\sqrt{\sum_{j=1}^{i}\beta_j + \sqrt{\left(\sum_{j=1}^{i}\beta_j\right)^2 + 4R_{\sigma,q}^2\sum_{j=i+1}^{|\mathcal{B}|}\sigma_j^2}} = e_{si,q} \tag{24}$$

$$\Leftrightarrow \Delta\sigma_{i,q} > 0, \forall i \in \mathcal{B}, \forall q \in \mathcal{R}.$$

*Proof:* Refer to Appendix B for a detailed proof.

*Proposition 3:* For the $i$th and $k$th ($i < k$) data brokers, the following two inequalities are satisfied when data brokers are both active data brokers and both inactive data brokers respectively, namely,

$$e_{si,q} > e_{sk,q}, \forall i,k \in \mathcal{B}_{o,q}, i < k, \forall q \in \mathcal{R}, \tag{25}$$

$$e_{si,q} \le e_{sk,q}, \forall i,k \in \mathcal{B}\setminus\mathcal{B}_{o,q}, i < k, \forall q \in \mathcal{R}. \tag{26}$$

*Proof:* Refer to Appendix C for a detailed proof, where an important lemma is introduced and also proven.

*Lemma:* Considering the non-negativity of $\sigma_j$ and $\beta_j$ for $\forall j \in \mathcal{B}$, the following inequality is satisfied:

$$e_k \le \sqrt{\frac{\sum_{j=i+1}^{k}\sigma_j^2}{\sum_{j=i+1}^{k}\beta_j}} \le e_i, \forall i,k \in \mathcal{B}, i < k. \tag{27}$$

Finally, the optimal uncertainty reduction of data brokers for the $q$th electricity retailer and the optimal total surplus for the $q$th electricity retailer and its data brokers are as follows. In other words, the maximum total surplus is the data profits that the $q$th electricity retailer and its data brokers can allocate. Facing passive end users, a positive optimal surplus for the coalition of data sharing can be expected:

$$\Delta\sigma_{i,q}^* = \begin{cases} \sigma_i - e_{si,q\min}\sqrt{\beta_i}, \forall i \in \mathcal{B}_{o,q} \\ 0 \qquad\qquad\quad, \forall i \in \mathcal{B}\setminus\mathcal{B}_{o,q} \end{cases}, \tag{28}$$

$$S_q^* = R_{\sigma,q}\left(\sigma_{e,q} - \sqrt{e_{si,q\min}^2\sum_{i\in\mathcal{B}_{o,q}}\beta_i + \sum_{i\in\mathcal{B}\setminus\mathcal{B}_{o,q}}\sigma_i^2}\right) \\ - \sum_{i\in\mathcal{B}_{o,q}}\beta_i\ln\frac{\sigma_i}{e_{si,q\min}\sqrt{\beta_i}}, \forall q \in \mathcal{R}. \tag{29}$$

### B. Asymmetric Nash Bargaining

In the coalition, electricity retailers and data brokers constitute a cooperative game. According to existing research, the Shapley value method and the nucleolus method are most often applied to allocate the benefits in the coalition of a cooperative game [33]. However, in these two methods, the basis of profit allocation is the marginal profits of coalition members. If electricity retailers are included in the coalition, the marginal profits of electricity retailers will contain the revenues from electricity transactions, which are not comparable to the marginal profits of data brokers that are from data sharing; if electricity retailers are excluded from the coalition, the results of data sharing are merely related to the marginal profit shares of data brokers, and it may be difficult to determine the proportion of profit allocation between electricity retailers and data brokers. More essentially, the limitation of the Shapley value method and nucleolus method is that they are almost only suitable for homogeneous coalition members.

As mentioned earlier, the contributions of electricity retailers and data brokers are heterogeneous. In data sharing, electricity retailers are the initial founders, while data brokers are the follow-up promoters because at the preliminary stage of the current data market, supply is driven by demand. Therefore, the asymmetric Nash bargaining model, which is suitable for profit allocation considering different types of contributors, is used. The key to the asymmetric Nash bargaining model is the marginal profit shares of each coalition member, which require that the sum be equal to one and that the definition of marginal profit shares be fair. Fortunately, the marginal profits of electricity retailers and data brokers to the coalition are numerically comparable: the marginal profit of each electricity retailer is its solution of optimal total surplus because this part of optimal total surplus will be zero without its presence in the coalition and its own revenues from electricity transactions are independent of data; the marginal profit of each data broker is the difference in the optimal total surplus of all electricity retailers that use its data services between its presence and absence in the coalition. The sum of marginal profit shares should be one; thus, each coalition member's marginal profit share should equal its marginal profit divided by the sum of marginal profits of all electricity retailers and data brokers, and the marginal profit shares for the $q$th electricity retailer and the $i$th data broker are

$$\tau_q^{er} = \frac{S_q^*}{\sum\limits_{q \in \mathcal{R}} S_q^* + \sum\limits_{\substack{i \in \bigcup\limits_{q \in \mathcal{R}} \mathcal{B}_{o,q}}} \sum\limits_{q \in \mathcal{R}} \left( S_q^* - S_{q,T}^* \right)}, \forall q \in \mathcal{R}, \quad (30)$$

$$\tau_i^{db} = \frac{\sum\limits_{q \in \mathcal{R}} \left( S_q^* - S_{q,T}^* \right)}{\sum\limits_{q \in \mathcal{R}} S_q^* + \sum\limits_{\substack{i \in \bigcup\limits_{q \in \mathcal{R}} \mathcal{B}_{o,q}}} \sum\limits_{q \in \mathcal{R}} \left( S_q^* - S_{q,T}^* \right)}, \forall i \in \bigcup\limits_{q \in \mathcal{R}} \mathcal{B}_{o,q}. \quad (31)$$

**Remark:** Note that the marginal profit shares should be nonnegative because they are the proportion of profit allocated to coalition members; otherwise, coalition members will receive negative allocated profits, which does not satisfy individual rationality and will lead to the split of the coalition. For the $q$th electricity retailer, the marginal profit shares are apparently nonnegative because the optimal surplus is zero when each $\Delta \sigma_{i,q}$ equals zero, and it will be nonnegative in the maximization problem regardless of the specific optimization solution. For data brokers, the conclusion that the criterion for the $i$th data broker $e_{si,q}$ will increase due to the absence of any active data broker can be intuitively drawn according to their expressions in Eq. (24) and the proof of **Proposition 1**, thus the cost-effectiveness index of the $i$th data broker $e_i$ will probably be lower than $e_{si,q}$ because $e_{si,q}$ remains unchanged. The higher criterion may allow some active data brokers to become inactive data brokers and completely lose their profits, while other active data brokers may reduce uncertainty less according to the optimal solution in Eq. (28) and receive a smaller profit allocation referring to the proof of **Proposition 2**, which makes the optimal surplus in the absence of one data broker less than that in its presence. Similarly, it can be derived that the criteria and the optimal surplus will not be affected between the presence and absence of any inactive data broker. In addition, from the perspective of the optimization model in Eq. (18) − (19), the absence of the $i$th data broker transforms the constraint "$0 \le \Delta \sigma_{i,q} < \sigma_i$" change into the constraint "$\Delta \sigma_{i,q} = 0$". It is obvious that the constraint "$\Delta \sigma_{i,q} = 0$" is the proper subset of the constraint "$0 \le \Delta \sigma_{i,q} < \sigma_i$"; thus, the feasible region of the surplus maximization problem decreases, and the optimal surplus will be no larger than that in the presence of the $i$th data broker. Therefore, the marginal profit shares of all coalition members are nonnegative.

Based on the marginal profit shares, the asymmetric Nash bargaining model is proposed for the profit allocation between electricity retailers and data brokers as follows, where the objective function can be regarded as a form of the utility function of the coalition.

$$\max_{D_q^{er}, D_i^{db}} U_C = \prod_{q \in \mathcal{R}} \left[ D_q^{er} - \left( p_{q,t} - \pi_{q,t} \right) \mathbb{E} X_{q,t} + R_{\sigma,q} \sigma_{e,q} \right]^{\tau_q^{er}}$$
$$\cdot \prod_{i \in \bigcup\limits_{q \in \mathcal{R}} \mathcal{B}_{o,q}} \left( D_i^{db} \right)^{\tau_i^{db}} \quad (32)$$

$$s.t. \quad \sum_{q \in \mathcal{R}} \left[ D_q^{er} - \left( p_{q,t} - \pi_{q,t} \right) \mathbb{E} X_{q,t} + R_{\sigma,q} \sigma_{e,q} \right]$$
$$+ \sum_{i \in \bigcup\limits_{q \in \mathcal{R}} \mathcal{B}_{o,q}} D_i^{db} \le \sum_{q \in \mathcal{R}} S_q^* \quad (33)$$

$$D_q^{er} - \left( p_{q,t} - \pi_{q,t} \right) \mathbb{E} X_{q,t} + R_{\sigma,q} \sigma_{e,q} \ge 0, \forall q \in \mathcal{R} \quad (34)$$

$$D_i^{db} \ge 0, \forall i \in \bigcup\limits_{q \in \mathcal{R}} \mathcal{B}_{o,q}. \quad (35)$$

To solve the optimization model, the logarithm of the utility function is used for the convenience of the subsequent derivation. The partial derivatives of the logarithm of the utility function with respect to the aggregate benefits of each coalition

---

**Calculation process 2:** Profit allocation using asymmetric Nash Bargaining (for electricity retailers)

1. **Input:** Given the uncertainty level $\sigma_i$ and privacy sensitivity $\beta_i$ of the groups of end users connected to data brokers, electricity retailers can directly caclculate the cost-effectiveness index $e_i$ of them.

2. **Assumptions:**
   a) The data collected by data brokers from their end users are irrelevant, thus the variances of different groups of end users are additive for electricity retailers.
   b) Electricity retailers and data brokers possess the symmetric information before IoT data sharing, *i.e.*, the load forecast accuracy of electricity retailers and data brokers on the same group of end users will be almost equal.

3. Maximize the total consumers' surplus related to electricity retailers and data brokers as the end users are regarded as passive participants in IoT data sharing.

4. Form the coalition of IoT data sharing between electricity retailers and data brokers whose cost-effectiveness index $e_i$ exceeds the cost-effectiveness standard of electricity retailers $e_{si,q}$ according to the results of consumers' surplus maximization.

5. Calculate the marginal contribution of each electricity retailer or data broker to the coalition of IoT data sharing.

6. Obtain the marginal contribution rates of each electricity retailer or data broker by using the marginal contribution of him divided by the total marginal contributions of all electricity retailers or all data brokers.

7. Share the total data profits fair between electricity retailers and data brokers based on the margin contribution rates through asymmetric Nash bargaining.

8. **Outputs:** profits received by electricity retailers $D_q^{er}$ and profit allocated to data brokers $D_i^{db}$.

member which contain its marginal profits and its revenues without data sharing are

$$\frac{\partial \ln U_C}{\partial D_q^{er}} = \frac{\tau_q^{er}}{D_q^{er} - \left(p_{q,t} - \pi_{q,t}\right)\mathbb{E}X_{q,t} + R_{\sigma,q}\sigma_{e,q}}, \forall q \in \mathcal{R}, \quad (36)$$

$$\frac{\partial \ln U_C}{\partial D_i^{db}} = \frac{\tau_i^{db}}{D_i^{db}}, \forall i \in \bigcup_{q \in \mathcal{R}} \mathcal{B}_{o,q}. \quad (37)$$

The optimal solution of the Nash bargaining model requires that the partial derivatives of each coalition member are equal; thus, the aggregate benefits of the $q$th electricity retailer and the $i$th data broker are derived as follows:

$$D_q^{er} = \tau_q^{er}\sum_{q \in \mathcal{R}} S_q^* + \left(p_{q,t} - \pi_{q,t}\right)\mathbb{E}X_{q,t} - R_{\sigma,q}\sigma_{e,q}, \forall q \in \mathcal{R}, \quad (38)$$

$$D_i^{db} = \tau_i^{db}\sum_{q \in \mathcal{R}} S_q^*, \forall i \in \bigcup_{q \in \mathcal{R}} \mathcal{B}_{o,q}. \quad (39)$$

Obviously, the results of profit allocation satisfy Pareto optimality, individual rationality, no exploitation, and budget balance [34]. The complete calculation processes of profit allocation can also be summarized.

## IV. Case Studies

### A. Basic Setup

In the case studies, the practical data used are mainly the load demand data of end users and the market price of data in electricity markets. To ensure the consistency and comparability of our research, the load demand data are smart meter data of the Commission for Electricity Regulation (CER) in Ireland, and the electricity wholesale market price data are from the PJM market in the US. In terms of the electricity retailer market price data, it is common for electricity retailers to provide the optimal pricing schemes. However, the impact of TOU prices on end users' electricity consumption in different time intervals and their choice on electricity retailers will not change the form of data valuation; hence, the flat retail price of 0.25 $/kWh is applied. Negative and positive penalty rates for electricity excess and deficit are set to 0.3 and 0.4 respectively to distinguish the results of different directions of electricity imbalances. Besides, to test the performance of IoT data sharing on calculation time, all programs are run on a personal computer with an Intel Core i7-8550U 1.80-GHz CPU and 16 GB of RAM.

### B. Results of Data Valuation

#### 1) Data Valuation

In our previous work [12], the definition of data value rates based on the same scenario as this paper is elaborated, and the formulas for the data value rates are derived under parametric and nonparametric estimation of the probability density function (PDF) of random variables. However, all candidate distributions applied in parametric estimation have disadvantages; for example, the cumulative distribution function (CDF) of the Gaussian distribution cannot be precisely derived in closed form, and the logistic distribution is not commonly used in distribution fitting. Therefore, the data value rate under kernel density estimation (KDE) is used to depict the distribution of load forecast errors.

The data value rate under KDE can be expressed as [12]

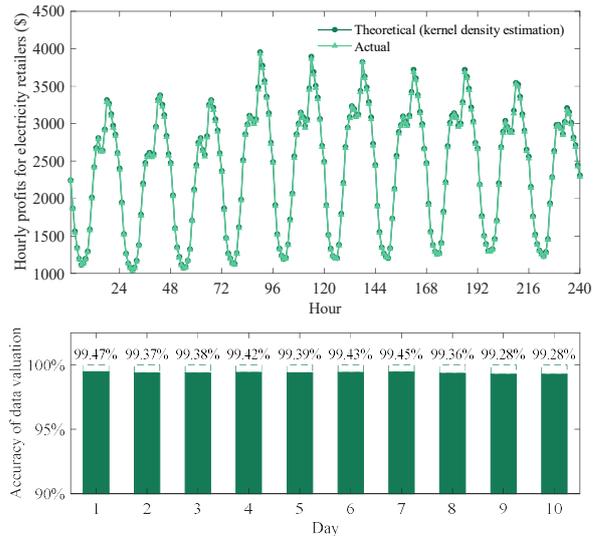

Fig. 5. Comparison of hourly theoretical and actual optimal profits for electricity retailers and accuracy of data value rates under KDE.

$$V_{\sigma,q} = \frac{\sum_{m=1}^{M}\left[h^* V_{\sigma,q}\left(\gamma_t^{m*}\right) + \left(\lambda_t^- - \lambda_t^+\right)\left(\gamma_t^* - \gamma_t^{m*}\right)x_t^m\right]}{M\sqrt{(h^*)^2 + \frac{1}{M}\sum_{m=1}^{M}\left(x_t^m - \frac{1}{M}\sum_{t=1}^{M}x_t^i\right)^2}}. \quad (40)$$

where $M$ is the number of kernel functions used in KDE, $h^*$ is the optimal bandwidth in KDE, and $\gamma_t^{m*}$ is the quantile of the optimal bid for electricity retailer $\tilde{X}_t^*$ in the $m$th kernel function.

The aggregate uncertainty that electricity retailers face $\sigma_e$ is assumed to be 0.1020. Based on the formula for the relationship between the optimal profits and the data value rate in Eq. (12), the data value rate under KDE in Eq. (40), and the assumed $\sigma_e$, the theoretical optimal profits under KDE can be calculated and compared with the actual results following the two-settlement market system. A 10-day example of hourly theoretical and actual optimal profits for electricity retailers through 1000 repeated tests is shown in Fig. 5, and the accuracy data of the data value rates under KDE in the 10 days are also listed.

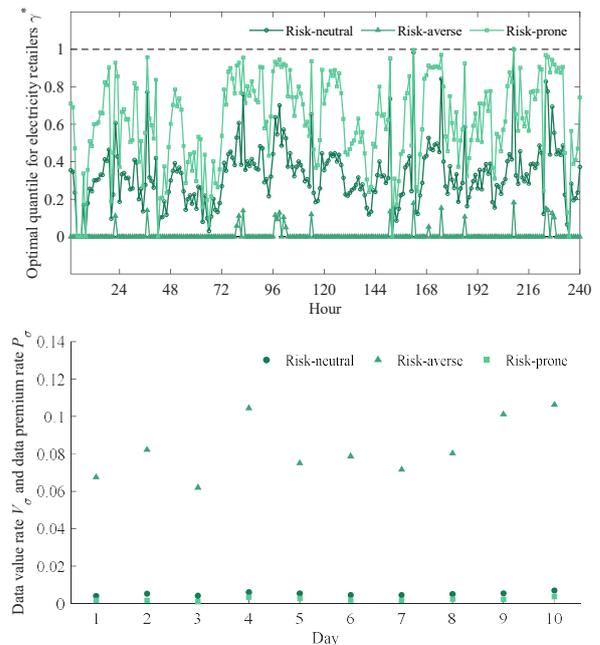

Fig. 6. Comparison of hourly optimal quantiles for electricity retailers with different risk preferences and the corresponding daily mean data value rates, risk-averse data premium rates, and risk-prone data premium rates.

Fig. 5 shows that the theoretical optimal profits are almost coincident with the actual optimal profits, and the accuracy of data value rates under KDE is extremely high, remaining above 99%. Considering the KDE's universality in load demand distribution fitting and high accuracy in data valuation, it can be used directly in the relevant derivation and application scenarios.

### 2) Data Valuation for Electricity Retailers with Different Risk Preferences

The effectiveness of data valuation under KDE is validated; thus, it is used to calculate the data value rate and data premium rate for electricity retailers with different risk preferences. To avoid the impact of different parameters on the results, the degree of risk aversion $\eta^a$ and the degree of risk proneness $\eta^p$ are both set to 1, and the conservative quantile $\varepsilon^a$ and the aggressiveness quantile $\varepsilon^p$ are both set to 0.1. A 10-day example of hourly optimal quantiles for electricity retailers with different risk preferences and the corresponding daily mean data value rates, risk-averse data premium rates, and risk-prone data premium rates through 1000 repeated tests are shown in Fig. 6.

Fig. 6 shows that the optimal quantiles of risk-averse, risk-neutral, and risk-prone electricity retailers strictly conform to the formulas in Eq. (10) – (11) and Fig. 3. The optimal quantiles of risk-averse electricity retailers are no larger than those of risk-neutral electricity retailers, while the optimal quantiles of risk-neutral electricity retailers are no larger than those of risk-prone electricity retailers. In addition, the differences in optimal quantiles between risk-neutral and risk-prone electricity retailers are usually narrower than those of the optimal quantiles between risk-averse and risk-neutral electricity retailers because of the feature of optimal solutions, which can also be found in Fig. 3 for a theoretical basis. Therefore, the above results can generally be used to explain that the risk-averse data premium rate is much larger than the risk-prone data premium rate at the same values of $\eta$ and $\varepsilon$.

### 3) Sensitivity Analysis for Data Premium Rates

From the discussion of Section II-C, it is known that the data value rate is related only to market price data, while the risk-averse and risk-prone data premium rates are not only related to market price data but depend on the degree of risk aversion $\eta^a$ or the degree of risk proneness $\eta^p$ and the conservative quantile $\varepsilon^a$ or the aggressiveness quantile $\varepsilon^p$. To de-termine the impact of these parameters to measure risk preferences on the data premium rates, a sensitivity analysis is employed, where the degree of risk aversion $\eta^a$ and the degree of risk proneness $\eta^p$ vary from 0.2 to 1.0 with an interval of 0.2, and the conservative quantile $\varepsilon^a$ and the aggressiveness quantile $\varepsilon^p$ vary from 0.05 to 0.25 with an interval of 0.05. The results of the sensitivity analysis are depicted in Fig. 7.

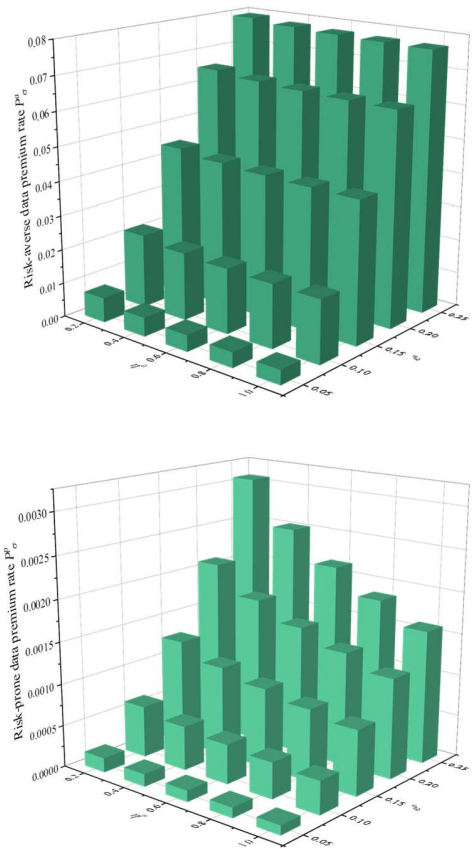

Fig. 7. Risk-averse and risk-prone data premium rates under the different degree of risk-aversion $\eta^a$ or degree of risk-proneness $\eta^p$ and the different conservative quantile $\varepsilon^a$ or aggressiveness quantile $\varepsilon^p$.

Fig. 7 shows that risk-averse and risk-prone data premium rates overall follow a similar pattern in which data premium rates decrease as the degree of risk aversion $\eta^a$ and the degree of risk proneness $\eta^p$ increase, and data premium rates increase as the conservative quantile $\varepsilon^a$ and the aggressiveness quantile $\varepsilon^p$ increase. It can be interpreted that as the degree of risk aversion $\eta^a$ or the degree of risk proneness $\eta^p$ increases, the weights of the CVaR terms are higher in the optimization models for risk-averse and risk-prone electricity retailers, and electricity retailers obtain optimal solutions that are further away from the optimal solutions of optimizing the profit expectations to en-sure the maximization of the sum of the profit expectations and the CVaR terms. Thus data premium rates will decrease from the perspective of profit expectations; as the conservative quantile $\varepsilon^a$ or the aggressiveness quantile $\varepsilon^p$ increase, more left-side or right-side tail risks are taken into consideration, and electricity retailers obtain optimal solutions that are closer to the optimal solutions of optimizing the profit expectations to fit with the when the changes that $\varepsilon^a$ and $1-\varepsilon^p$ are approaching the opposite sides, and thus data premium rates will increase.

TABLE I
PARAMETERS OF TEN DATA BROKERS

| Parameter | Data brokers | | | | | | | | | |
|---|---|---|---|---|---|---|---|---|---|---|
| | #1 | #2 | #3 | #4 | #5 | #6 | #7 | #8 | #9 | #10 |
| Uncertainty level $\sigma_i$ | 0.0229 | 0.0315 | 0.0464 | 0.0354 | 0.0422 | 0.0372 | 0.0317 | 0.0314 | 0.0155 | 0.0087 |
| Privacy sensitivity $\beta_i$ | 0.0216 | 0.0421 | 0.1426 | 0.0895 | 0.1483 | 0.1289 | 0.1033 | 0.1411 | 0.0445 | 0.0210 |
| Cost-effectiveness index $e_i$ | 0.1558 | 0.1535 | 0.1229 | 0.1183 | 0.1096 | 0.1036 | 0.0986 | 0.0836 | 0.0735 | 0.0600 |

### C. Results of Profit Allocation

#### 1) Parameters of Data Brokers

To cover more situations of data brokers, ten data brokers are assigned to participate in the data sharing and their parameters including the uncertainty level $\sigma_i$, privacy sensitivity $\beta_i$, and cost-effectiveness index $e_i$, are differently given and the specific values are shown in Table I. Note that the ten data brokers are arranged in descending order of $e_i$ in line with the assumption in Eq. (26).

#### 2) Data Brokers in Surplus Maximization

In the profit allocation, electricity retailers should maximize the total surplus of the coalition in advance. Based on the formulas for data brokers' cost-effectiveness indices $e_i$ in Eq. (25) and criteria to be active data brokers $e_{si,q}$ in Eq. (29), an example of the optimal solutions of data brokers' uncertainty reduction is shown in Fig. 8, where the comparison of their $e_i$ and $e_{si,q}$ are also shown to validate the derivation in **Proposition 2** and Eq. (33).

Fig. 8 shows that there are seven active data brokers whose cost-effectiveness indices $e_i$ exceed their cost-effectiveness criteria $e_{si,q}$ and three inactive data brokers whose cost-effectiveness indices $e_i$ are lower than their cost-effectiveness criteria $e_{si,q}$ at the given marginal revenue of the electricity retailer. Logically, only the seven active data brokers' data are cost-effective enough for the electricity retailer and they achieve corresponding uncertainty reduction as the optimal solutions indicate, while the other three inactive data brokers do not have the chance to share their data with the electricity retailer. Note that there are differences among the cost-effectiveness of the seven active data brokers' data; thus, the data broker with higher $e_i$ can reduce a higher proportion of the uncertainty reduction it can achieve. In addition, the different monotonicity of cost-effectiveness criteria $e_{si,q}$ for active and inactive data brokers described in **Proposition 3** is also validated.

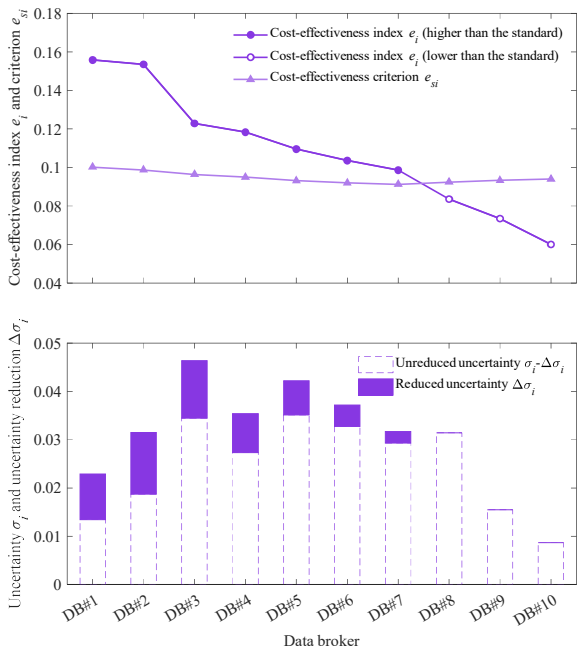

Fig. 8. An example of cost-effectiveness indices $e_i$ and criteria to be active $e_{si,q}$ of data brokers, and corresponding optimal solutions of their uncertainty reduction.

#### 3) Case 1: One Risk-Neutral Electricity Retailer

For the simplicity of profit allocation, the case contains one risk-neutral electricity retailer, and ten data brokers are considered first. The data value rates of the electricity retailer are calculated based on KDE, and the electricity transactions and profit allocation are simulated in the time span of half a year. Then, the total marginal profits in half a year and the corresponding marginal profit shares are depicted in Fig. 9.

Fig. 9 shows that the electricity retailer contributes the most marginal profits among all coalition members under the Nash

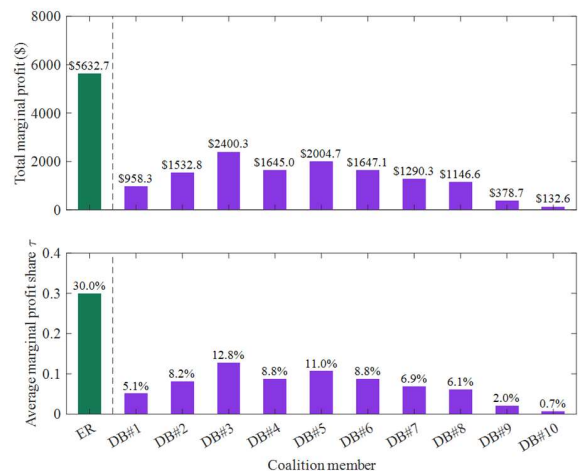

Fig. 9. The total marginal profits in half a year and the corresponding marginal profit shares of one risk-neutral electricity retailer and ten data brokers.

bargaining model, which is consistent with our analysis in

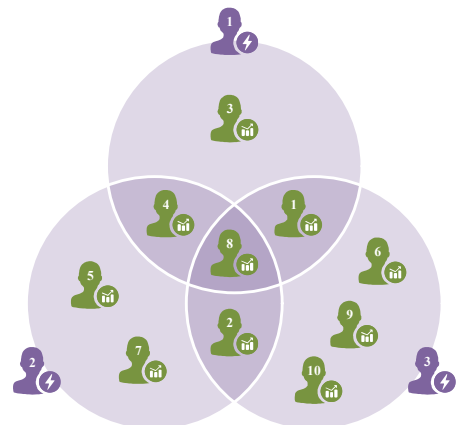

Fig. 10. An example of the coalition of data sharing comprises of three electricity retailers and ten data brokers.

Section III-B, while data brokers also contribute their distinct marginal profits based on their optimal uncertainty reduction solved through the surplus maximization problem. Finally, the electricity retailer and data brokers should allocate the total surplus $5,632.7 by marginal profit shares.

For the profit enhancement of the electricity retailer and data brokers, the electricity retailer who are responsible for 6,064 end users can receive approximately $8,132,926.3 as the total revenues in half a year. In IoT data sharing, the demand of uncertainty reduction for the electricity retailer bring $18,769.0 as additional benefits to the electricity retailer and data brokers, which is 2.3‰ of the total revenues. Through profit allocation, $5,632.7 in the total revenues are allocated to the electricity retailer, which is 0.7‰ of his total revenues, and the rest $13,136.3 are allocated to data brokers, which are almost net profits for them. Although the additional benefits from IoT data sharing is relatively slim for the electricity retailer, this is only the benefits of the electricity retailer with high load forecasting accuracy in the aspect of further improving his load forecasting accuracy. If the electricity retailer utilizes IoT data sharing in other aspects of his business, he will obtain more additional profits and benefit more data brokers.

### 4) Case 2: Three Risk-Neutral Electricity Retailers

Considering that the coalition members should comprise several electricity retailers and several data brokers to fit reality,

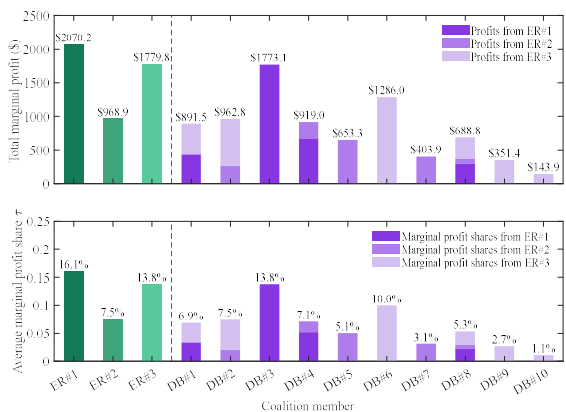

Fig. 11. The total marginal profits in half a year and the corresponding marginal profit shares of three risk-neutral electricity retailers and ten data brokers.

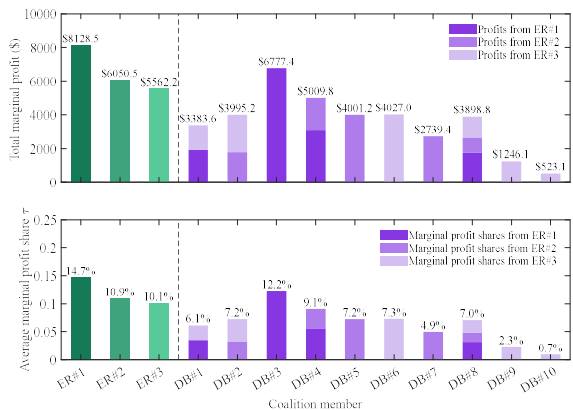

Fig. 12. The total marginal profits in half a year and the corresponding marginal profit shares of three risk-averse or risk-prone electricity retailers and ten data brokers.

another case containing three risk-neutral electricity retailers and ten data brokers is proposed, where the three electricity retailers share data with their specific group data brokers and one data broker may also serve multiple electricity retailers, as they have more data sources of end users. The data sharing coalition is shown in Fig. 10, where three electricity retailers obtain data from 4, 5, and 6 data brokers, and data brokers (DB) #1, #2, #4, and #8 serve multiple electricity retailers. Similarly, the profit allocation is simulated in the time span of half a year, and the total marginal profits in half a year and the corresponding marginal profit shares of electricity retailers and data brokers are depicted in Fig. 11.

Fig. 11 shows that the three electricity retailers together still contribute the most marginal profits among all coalition members, but the aggregate marginal profits of the three electricity retailers, which are also the total surplus that can be allocated to all coalition members, decrease to $4,818.9. The reduction in total surplus is because, with the dispersion of data brokers, they can only achieve lower uncertainty reduction for different electricity retailers they serve since their end users' privacy sensitivity is fixed.

### 5) Case 3: Three Risk-Averse or Risk-Prone Electricity Retailers

Electricity retailers can have risk preferences other than risk neutrality, which may affect their marginal revenues from data according to the derivation in Section II-C. To simulate electricity retailers with different risk preferences, the coalition members and their relationships between them are the same as those in Case 2, and the three electricity retailers are a risk-averse electricity retailer with $\eta^a = 0.2$ and $\varepsilon^a = 0.1$, a risk-prone electricity retailer with $\eta^p = 2$ and $\varepsilon^p = 0.1$, and a risk-averse electricity retailer with $\eta^a = 0.1$ and $\varepsilon^a = 0.05$. Similarly, the profit allocation is simulated in the time span of half a year, and the total marginal profits in half a year and the corresponding marginal profit shares of electricity retailers and data brokers are depicted in Fig. 12.

Fig. 12 indicates that the marginal profit shares of the three electricity retailers change sharply due to their distinct risk preferences. In addition, the aggregate marginal profits of the three electricity retailers increase to $19,741.2 because the data premium rates are always much larger than the data value rates.

### D. Discussions

#### 1) Comparison of profit allocation among different cases

Based on the three cases mentioned above, the profit allocation results and proportions among those cases can be compared in Fig. 13.

Fig. 13 shows that it is better for data brokers to serve an electricity retailer with many end users, as the total surplus and the profit allocation proportion are both higher when they cannot judge the electricity retailer's risk preferences; otherwise, they should choose electricity retailers with extreme risk preferences whose profit allocation proportion is lower, but the total surplus will be sufficient to ensure that data brokers receive more profits.

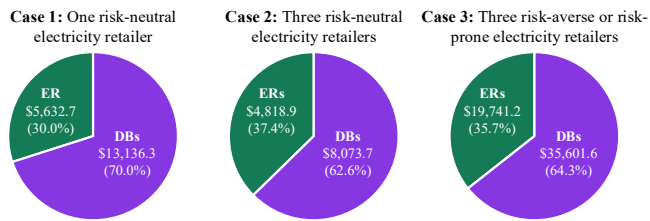

Fig. 13. Comparison of the profit allocation results and proportions among three cases.

### 2) Comparison of profit allocation among different methods

To compare different profit allocation methods, the commonly-used Shapley value method and the profit allocation method proposed in [17] are regarded as the benchmarks. Using Case 1 above as the case, the profit allocation proportions using can be compared in Fig. 14.

Fig. 14 shows that the electricity retailer cannot receive any additional profits from IoT data sharing when using Shapley value method. In Shapley value method, the allocated profit is average marginal contribution of each coalition member under all possible orders of coalition joining, thus in the scenario of electricity retail, all coalition member will not receive any profits before the electricity retailers join the coalition. The marginal contributions of electricity retailers in all possible orders will be considered, while the marginal contributions of data brokers in part of possible orders will be considered. To avoid the problem mentioned above, electricity retailers should be considered as permanent members in the coalition of IoT data sharing, and Shapley value method is merely used for data brokers. Under such circumstance, it is hard to be fair when using Shapley value method to allocate profits in IoT data sharing. Moreover, the proportion for the electricity retailer is also somewhat unfair when using the profit allocation method proposed in [17], because the electricity retailer is the main initiator of IoT data sharing and he provides the opportunities of obtaining data value to data brokers, thus only 4.6% of total profits are allocated to the electricity retailer is unreasonable. By comparison, the profit allocation method using asymmetric Nash bargaining is the most reasonable and fair one for both the electricity retailer and data brokers, where their heterogeneous contributions are comprehensively considered. The electricity retailer is in a crucial position and he receives relatively high 30.0% of profits, while all data brokers receive the rest 70.0% of profits directly related to their margin contributions.

### 3) Calculation time

To demonstrate whether the calculation time of data valuation and profit allocation is acceptable for the application of IoT data sharing in electricity transactions, the time needed for hourly calculation of data valuation and profit allocation programs is calculation by averaging the time of one thousand times of program runs.

At time intervals of one hour and still using Case 1 above as the case, data valuation needs 0.0034 seconds and profit allocation needs 0.2560 seconds. For comparison, profit allocation using Shapley value method needs 22.1912 seconds, profit allocation using nucleolus method needs 22.0024 seconds. It can be learned from formulas that the calculation time of profit allocation using Shapley value method or nucleolus method will exhibit a factorial growth rate as the scale of end users increases, while the calculation time of profit allocation using asymmetric Nash bargaining only exhibit a linear growth rate as the scale of end users increases. Therefore, profit allocation using asymmetric Nash bargaining can even meet the application requirements of 5-minute ultrashort-term load forecasting, while profit allocation using Shapley value method or nucleolus method may not have satisfactory computational efficiency, especially when the scale of electricity retailers, data brokers, and end users are large.

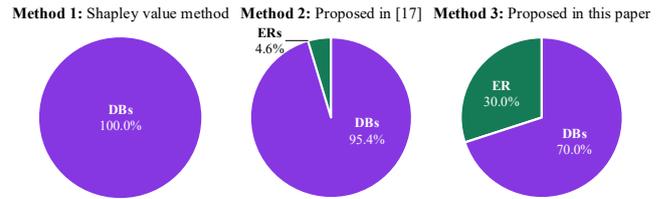

Fig. 14. Comparison of the profit allocation proportions among three profit allocation methods.

## V. CONCLUSIONS AND FUTURE WORK

Focusing on the flaws in IoT data sharing, the models in the two main stages, including data valuation and profit allocation, are remoulded in this paper. For the data valuation, a comprehensive model incorporating the impacts of feature attributes of data buyers on the data value is proposed to reflect the fact that data value is related to the demanders' use; for the profit allocation, an asymmetric Nash bargaining model is introduced considering the heterogeneous contributions of data buyers and sellers to the coalition of data sharing. The scenario in which electricity retailers participate in electricity transactions is applied for practical analysis, where the feature attributes are considered the risk preferences of electricity retailers and the profit allocation is between electricity retailers and data brokers who collect end users' data and provide private data services. The results in practical cases validate that the data premium rates caused by risk preferences will be added to data value rates and become an important part of the marginal revenues of electricity retailers. Moreover, the profit allocation between electricity retailers and data brokers is reasonable because electricity retailers obtain more profits due to their contributions to determining the data demand and promoting the formation of data sharing, while data brokers can also be inspired in choosing electricity retailers.

Future works can be extended from several interesting directions in the field of IoT data sharing. First is the generalization of application scenarios. IoT data are valuable intangible assets that can be broadly used in different practical scenarios, and their value is not limited to uncertainty reduction. The second is the IoT data audit. Unlike commodities, IoT data more or less imply the privacy of entities associated with the data. To avoid privacy not being properly exposed in the application, the regulator needs to audit the target and content of data sharing. The third is the mechanism design of IoT data markets. With the improvement of regulation, data will not be monopolized by some statistical departments or data platforms but will be towards free circulation. Hence, sound trading and settlement mechanisms need to be created to guarantee the efficiency of IoT data market.

## APPENDIX

### A. Proof of Proposition 1

For the relationship in Eq. (22), it can be proven through the reduction to absurdity. It is assumed that there exists $i$ that satisfies $i < k$ and that the $i$th data broker is an inactive data broker, thus "$\Delta\sigma_{i,q} = 0$" is part of the optimal solution, while "$\Delta\sigma_{k,q} > 0$" should also be satisfied as the $k$th data broker is an active data broker.

The first-order partial derivative of the total surplus $S_q$ with respect to the uncertainty reduction of the $i$th and $k$th data brokers for the $q$th electricity retailer $\Delta\sigma_{i,q}$ and $\Delta\sigma_{k,q}$ based on the conditions mentioned above can be obtained as

$$\frac{\partial S_q}{\partial \Delta\sigma_{i,q}} = \frac{\sigma_i^2 R_{\sigma,q} - \beta_i \sqrt{\sum_{i\in\mathcal{B}}\left(\sigma_i - \Delta\sigma_{i,q}\right)^2}}{\sigma_i \sqrt{\sum_{i\in\mathcal{B}}\left(\sigma_i - \Delta\sigma_{i,q}\right)^2}}, \quad (A.1)$$

$$\frac{\partial S_q}{\partial \Delta\sigma_{k,q}} = \frac{\left(\sigma_k - \Delta\sigma_{k,q}\right)^2 R_{\sigma,q} - \beta_k \sqrt{\sum_{i\in\mathcal{B}}\left(\sigma_i - \Delta\sigma_{i,q}\right)^2}}{\left(\sigma_k - \Delta\sigma_{k,q}\right)\sqrt{\sum_{i\in\mathcal{B}}\left(\sigma_i - \Delta\sigma_{i,q}\right)^2}}. \quad (A.2)$$

Considering that the $k$th data broker is an active data broker with a positive $\Delta\sigma_{k,q}$, its first-order partial derivative should equal zero, namely,

$$\frac{\left(\sigma_k - \Delta\sigma_{k,q}\right)^2}{\beta_k} = \frac{\sqrt{\sum_{i\in\mathcal{B}}\left(\sigma_i - \Delta\sigma_{i,q}\right)^2}}{R_{\sigma,q}}. \quad (A.3)$$

However, based on the assumption in Eq. (21), we have

$$e_i \geq e_k = \frac{\sigma_k^2}{\beta_k} > \frac{\left(\sigma_k - \Delta\sigma_{k,q}\right)^2}{\beta_k} = \frac{\sqrt{\sum_{i\in\mathcal{B}}\left(\sigma_i - \Delta\sigma_{i,q}\right)^2}}{R_{\sigma,q}}. \quad (A.4)$$

Eq. (A.4) indicates that

$$\left.\frac{\partial S_q}{\partial \Delta\sigma_{i,q}}\right|_{\Delta\sigma_{i,q}=0} = \frac{\sigma_i^2 R_{\sigma,q} - \beta_i \sqrt{\sum_{i\in\mathcal{B}}\left(\sigma_i - \Delta\sigma_{i,q}\right)^2}}{\sigma_i \sqrt{\sum_{i\in\mathcal{B}}\left(\sigma_i - \Delta\sigma_{i,q}\right)^2}} > 0. \quad (A.5)$$

Therefore, based on the continuous derivability of the total surplus $S_q$ at "$\Delta\sigma_{i,q} = 0$", a small positive $\Delta\sigma_{i,q}$ can always be found to increase the optimal total surplus, which contradicts the assumption that the $i$th data broker is an inactive data broker, that is, the proposition that "$\Delta\sigma_{i,q} = 0$" is part of the optimal solution will be falsified.

Hence, if the $k$th data broker is an active data broker, then any $i$th data broker with a smaller serial number ($i < k$) will be an active data broker. Similarly, the relationship in Eq. (23) can also be proven through the reduction to absurdity. ∎

### B. Proof of Proposition 2

To solve the surplus maximization model in Eq. (18) – (19), the first-order optimality condition is still used first, thus the first-order partial derivative of the total surplus $S_q$ with respect to the uncertainty reduction of the $i$th data broker for the $q$th electricity retailer $\Delta\sigma_{i,q}$ is

$$\frac{\partial S_q}{\partial \Delta\sigma_{i,q}} = -R_{\sigma,q}\frac{\partial\sqrt{\sum_{i\in\mathcal{B}}\left(\sigma_i - \Delta\sigma_{i,q}\right)^2}}{\partial \Delta\sigma_{i,q}} + \beta_i\frac{\mathrm{d}\ln\left(\sigma_i - \Delta\sigma_{i,q}\right)}{\mathrm{d}\Delta\sigma_{i,q}}$$

$$= R_{\sigma,q}\frac{\left(\sigma_i - \Delta\sigma_{i,q}\right)}{\sqrt{\sum_{i\in\mathcal{B}}\left(\sigma_i - \Delta\sigma_{i,q}\right)^2}} - \beta_i\frac{1}{\sigma_i - \Delta\sigma_{i,q}}. \quad (B.1)$$

Let the first-order partial derivative equal to zero, and the condition can be transformed to

$$\left(\sigma_i - \Delta\sigma_{i,q}\right)^2 = \frac{\beta_i}{R_{\sigma,q}}\sqrt{\sum_{i\in\mathcal{B}}\left(\sigma_i - \Delta\sigma_{i,q}\right)^2}. \quad (B.2)$$

However, there may be some data brokers that cannot reach their optimal solutions related to the first-order optimal conditions; thus, their $\Delta\sigma_{i,q}$ should be located at the boundary of the feasible intervals, namely, zero. We define set $\mathcal{B}$ as composed of all data brokers and set $\mathcal{B}_{o,q}$ as composed of data brokers whose optimal solution is not zero for the $q$th electricity retailer. Data brokers whose optimal solution is not zero are also called active data brokers, and other data brokers are called inactive data brokers. The conditions of active data brokers are summarized as follows:

$$\sum_{i\in\mathcal{B}_{o,q}}\left(\sigma_i - \Delta\sigma_{i,q}\right)^2 = \sum_{i\in\mathcal{B}_{o,q}}\frac{\beta_i}{R_{\sigma,q}}\sqrt{\sum_{i\in\mathcal{B}}\left(\sigma_i - \Delta\sigma_{i,q}\right)^2}$$

$$= \frac{\sum_{i\in\mathcal{B}_{o,q}}\beta_i}{R_{\sigma,q}}\sqrt{\sum_{i\in\mathcal{B}_{o,q}}\left(\sigma_i - \Delta\sigma_{i,q}\right)^2 + \sum_{i\in\mathcal{B}\backslash\mathcal{B}_{o,q}}\sigma_i^2}. \quad (B.3)$$

Through the derivation, the optimal solution for "$\Delta\sigma_{i,q} > 0$" is equivalent to the following inequality:

$$e_i = \frac{\sigma_i}{\sqrt{\beta_i}} > \frac{\sigma_i - \Delta\sigma_{i,q}}{\sqrt{\beta_i}}$$

$$= \frac{1}{\sqrt{2}R_{\sigma,q}}\sqrt{\sum_{i\in\mathcal{B}_{o,q}}\beta_i + \sqrt{\left(\sum_{i\in\mathcal{B}_{o,q}}\beta_i\right)^2 + 4R_{\sigma,q}^2\sum_{i\in\mathcal{B}\backslash\mathcal{B}_{o,q}}\sigma_i^2}}, \quad (B.4)$$

where set $\mathcal{B}_{o,q}$ is not a specific set but a set of active data suppliers including the $i$th data broker. From another perspective, the cost-effectiveness index $e_i$ of the $i$th data broker should satisfy Eq. (B.4) as long as he is an active data broker. According to the proof of *Proposition 1*, it is known that the smallest set of active data brokers that includes the $i$th data broker is the set that only contains data brokers with serial numbers from 1 to $i$, thus the cost-effectiveness index $e_i$ of the $i$th data broker must satisfy:

$$e_i > \frac{1}{\sqrt{2}R_{\sigma,q}}\sqrt{\sum_{j=1}^{i}\beta_j + \sqrt{\left(\sum_{j=1}^{i}\beta_j\right)^2 + 4R_{\sigma,q}^2\sum_{j=i+1}^{|\mathcal{B}|}\sigma_j^2}}. \quad (B.5)$$

Therefore, Eq. (24) is the necessary condition for the $i$th data broker to be an active data broker. Then, we will prove Eq. (24) is the sufficient condition for the $i$th data broker to be an active data broker, and the reduction to absurdity is used again. It is assumed that Eq. (24) is satisfied, but the $i$th data broker is an inactive data broker. According to Eq. (24) and the equivalence between Eq. (B.4) and the $i$th data broker being an active data broker, we have

$$e_i > \frac{\sqrt{\sum_{j=1}^{i}\beta_j + \sqrt{\left(\sum_{j=1}^{i}\beta_j\right)^2 + 4R_{\sigma,q}^2\sum_{j=i+1}^{|\mathcal{B}|}\sigma_j^2}}}{\sqrt{2}R_{\sigma,q}}, \quad (B.6)$$

$$e_i \leq \frac{\sqrt{\sum_{i\in\mathcal{B}_{o,q}}\beta_i + \sqrt{\left(\sum_{i\in\mathcal{B}_{o,q}}\beta_i\right)^2 + 4R_{\sigma,q}^2\sum_{i\in\mathcal{B}\setminus\mathcal{B}_{o,q}}\sigma_i^2}}}{\sqrt{2}R_{\sigma,q}}. \quad (B.7)$$

Because of **Proposition 1**, we know that set $\mathcal{B}_{o,q}$ only contains those active data brokers who have serial numbers smaller than $i$; thus, we assume that the maximum serial number of active data brokers is $k$ ($k < i$). Then, we have

$$\frac{\sqrt{\sum_{j=1}^{i}\beta_j + \sqrt{\left(\sum_{j=1}^{i}\beta_j\right)^2 + 4R_{\sigma,q}^2\sum_{j=i+1}^{|\mathcal{B}|}\sigma_j^2}}}{\sqrt{2}R_{\sigma,q}}$$
$$< \frac{\sqrt{\sum_{j=1}^{k}\beta_j + \sqrt{\left(\sum_{j=1}^{k}\beta_j\right)^2 + 4R_{\sigma,q}^2\sum_{j=k+1}^{|\mathcal{B}|}\sigma_j^2}}}{\sqrt{2}R_{\sigma,q}}, \forall k \in \mathcal{B}_{o,q}, k < i. \quad (B.8)$$

Through the derivation, we have the following inequality when $k = i - 1$:

$$\frac{\sqrt{\sum_{j=1}^{k}\beta_j + \sqrt{\left(\sum_{j=1}^{k}\beta_j\right)^2 + 4R_{\sigma,q}^2\sum_{j=k+1}^{|\mathcal{B}|}\sigma_j^2}}}{\sqrt{2}R_{\sigma,q}} < \sqrt{\frac{\sum_{j=k+1}^{i}\sigma_j^2}{\sum_{j=k+1}^{i}\beta_j}} = e_i. \quad (B.9)$$

According to the assumption, it is obvious that any possible set $\mathcal{B}_{o,q}$ should satisfy Eq. (B.6) – (B.7), but set $\mathcal{B}_{o,q}$ satisfies Eq. (B.9) when $k = i - 1$ based on our derivation:

$$e_i > \frac{1}{\sqrt{2}R_{\sigma,q}}\sqrt{\sum_{j=1}^{k}\beta_j + \sqrt{\left(\sum_{j=1}^{k}\beta_j\right)^2 + 4R_{\sigma,q}^2\sum_{j=k+1}^{|\mathcal{B}|}\sigma_j^2}}$$
$$= \frac{1}{\sqrt{2}R_{\sigma,q}}\sqrt{\sum_{i\in\mathcal{B}_{o,q}}\beta_i + \sqrt{\left(\sum_{i\in\mathcal{B}_{o,q}}\beta_i\right)^2 + 4R_{\sigma,q}^2\sum_{i\in\mathcal{B}\setminus\mathcal{B}_{o,q}}\sigma_i^2}}. \quad (B.10)$$

Note that Eq. (B.10) contradicts the assumption, which indicates that the assumption is falsified and Eq. (24) is the sufficient condition for the $i$th data broker to be an active data broker. In summary, Eq. (24) is the sufficient necessary condition for the $i$th data broker to be an active data broker. ∎

### C. Proof of Proposition 3

To prove **Proposition 3**, **Lemma** should be proved first.

According to the definition of $e_i$ in Eq. (20) and the basic assumption that the sequence $\{e_i\}$ is monotonically not increasing in Eq. (21), the right-side inequality in Eq. (27) can be proven as follows:

$$e_i = \sqrt{\overline{e_i^2}} = \sqrt{\frac{\sum_{j=i+1}^{k}e_i^2\beta_j}{\sum_{j=i+1}^{k}\beta_j}} \geq \sqrt{\frac{\sum_{j=i+1}^{k}e_j^2\beta_j}{\sum_{j=i+1}^{k}\beta_j}} = \sqrt{\frac{\sum_{j=i+1}^{k}\sigma_j^2}{\sum_{j=i+1}^{k}\beta_j}}. \quad (C.1)$$

Similarly, the left-side inequality in Eq. (27) can also be proven. Given that **Lemma** has been proved, for Eq. (25) in **Proposition 3**, because the $k$th data broker is an active data broker in Eq. (25), thus we can combine the **Lemma** and the cost-effectiveness criterion and then obtain

$$\sqrt{\frac{\sum_{j=i+1}^{k}\sigma_j^2}{\sum_{j=i+1}^{k}\beta_j}} \geq e_k \geq \frac{\sqrt{\sum_{j=1}^{k}\beta_j + \sqrt{\left(\sum_{j=1}^{k}\beta_j\right)^2 + 4R_{\sigma,q}^2\sum_{j=k+1}^{N}\sigma_j^2}}}{\sqrt{2}V_\sigma}. \quad (C.2)$$

Shifting the denominator of the right side to the left side and simultaneously squaring both sides, Eq. (C.2) is transformed to

$$\frac{2R_{\sigma,q}^2\sum_{j=i+1}^{k}\sigma_j^2}{\sum_{j=i+1}^{k}\beta_j} > \sum_{j=1}^{k}\beta_j + \sqrt{\left(\sum_{j=1}^{k}\beta_j\right)^2 + 4R_{\sigma,q}^2\sum_{j=k+1}^{N}\sigma_j^2}. \quad (C.3)$$

Shifting the first term of the right side to the left side and simultaneously squaring the nonnegative terms on both sides again, Eq. (C.3) is transformed to

$$\left(\frac{2R_{\sigma,q}^2\sum_{j=i+1}^{k}\sigma_j^2}{\sum_{j=i+1}^{k}\beta_j} - \sum_{j=1}^{k}\beta_j\right)^2 > \left(\sum_{j=1}^{k}\beta_j\right)^2 + 4R_{\sigma,q}^2\sum_{j=k+1}^{N}\sigma_j^2. \quad (C.4)$$

Expanding the left side and eliminating the same terms on both sides, Eq. (C.4) is transformed to

$$\left(\frac{2R_{\sigma,q}^2\sum_{j=i+1}^{k}\sigma_j^2}{\sum_{j=i+1}^{k}\beta_j}\right)^2 > 4R_{\sigma,q}^2\left(\sum_{j=k+1}^{N}\sigma_j^2 + \sum_{j=i+1}^{k}\sigma_j^2\frac{\sum_{j=1}^{k}\beta_j}{\sum_{j=i+1}^{k}\beta_j}\right). \quad (C.5)$$

Shifting the denominator of the left side to the right side, Eq. (C.5) is transformed to

$$4R_{\sigma,q}^4\left(\sum_{j=i+1}^{k}\sigma_j^2\right)^2 > 4R_{\sigma,q}^2\sum_{j=i+1}^{N}\sigma_j^2\sum_{j=1}^{k}\beta_j\sum_{j=i+1}^{k}\beta_j$$
$$- 4R_{\sigma,q}^2\sum_{j=k+1}^{N}\sigma_j^2\sum_{j=1}^{i}\beta_j\sum_{j=i+1}^{k}\beta_j. \quad (C.6)$$

Doubling both sides, adding terms to make up quadratic terms and extracting the square root on both sides, then

$$2R_{\sigma,q}^2\sum_{j=i+1}^{N}\sigma_j^2 + 2R_{\sigma,q}^2\sum_{j=k+1}^{N}\sigma_j^2 + \sum_{j=1}^{k}\beta_j\sum_{j=1}^{i}\beta_j$$
$$> \sqrt{\left(\sum_{j=1}^{k}\beta_j\right)^2 + 4R_{\sigma,q}^2\sum_{j=i+1}^{N}\sigma_j^2}\sqrt{\left(\sum_{j=1}^{k}\beta_j\right)^2 + 4R_{\sigma,q}^2\sum_{j=k+1}^{N}\sigma_j^2}. \quad (C.7)$$

Adding terms to make up quadratic terms and extracting the square root on both sides, then

$$\sqrt{\left(\sum_{j=1}^{i}\beta_j\right)^2 + 4R_{\sigma,q}^2\sum_{j=i+1}^{N}\sigma_j^2} - \sqrt{\left(\sum_{j=1}^{k}\beta_j\right)^2 + 4R_{\sigma,q}^2\sum_{j=k+1}^{N}\sigma_j^2}$$
$$> \sum_{j=1}^{k}\beta_j - \sum_{j=1}^{i}\beta_j. \tag{C.8}$$

Finally, Eq. (25) can be directly obtained through term shifting from Eq. (C.8). In terms of the proof of Eq. (26), the derivation should follow the steps from Eq. (C.2) – (C.8) and start with Eq. (C.9) as follows:

$$\sqrt{\frac{\sum_{j=i+1}^{k}\sigma_j^2}{\sum_{j=i+1}^{k}\beta_j}} \le \frac{1}{\sqrt{2}R_{\sigma,q}}\sqrt{\sum_{j=1}^{i}\beta_j + \sqrt{\left(\sum_{j=1}^{i}\beta_j\right)^2 + 4R_{\sigma,q}^2\sum_{j=i+1}^{N}\sigma_j^2}}. \tag{C.9}$$

In addition, it is also found that Eq. (C.10) can be proved based on **Proposition 2**.

$$e_{si,q} > e_{si+1,q}, i \in \mathcal{B}_{o,q}, i+1 \in \mathcal{B}\setminus\mathcal{B}_{o,q}, \forall q \in \mathcal{R}. \tag{C.10}$$ ∎

## REFERENCES

[1] Xinhua, "Guideline on Improving the Market-Based Allocation Mechanism of Production Factors in a Bid." Apr. 2020. [Online]. Available: http://english.www.gov.cn/policies

[2] C.-M. Yu, C.-F. Chien, and C.-J. Kuo, "Exploit the Value of Production Data to Discover Opportunities for Saving Power Consumption of Production Tools," *IEEE Trans. Semicond. Manuf.*, vol. 30, no. 4, pp. 345–350, Nov. 2017, doi: 10.1109/TSM.2017.2750712.

[3] J. Tastu, P. Pinson, P.-J. Trombe, and H. Madsen, "Probabilistic Forecasts of Wind Power Generation Accounting for Geographically Dispersed Information," *IEEE Trans. Smart Grid*, vol. 5, no. 1, pp. 480–489, Jan. 2014, doi: 10.1109/TSG.2013.2277585.

[4] C. Gilbert, J. Browell, and D. McMillan, "Leveraging Turbine-Level Data for Improved Probabilistic Wind Power Forecasting," *IEEE Trans. Sustain. Energy*, vol. 11, no. 3, pp. 1152–1160, Jul. 2020, doi: 10.1109/TSTE.2019.2920085.

[5] S. Buhan, Y. Özkazanç, and I. Çadırcı, "Wind Pattern Recognition and Reference Wind Mast Data Correlations With NWP for Improved Wind-Electric Power Forecasts," *IEEE Trans. Ind. Inform.*, vol. 12, no. 3, pp. 991–1004, Jun. 2016, doi: 10.1109/TII.2016.2543004.

[6] Y. Wang, C.-F. Chen, P.-Y. Kong, H. Li, and Q. Wen, "A Cyber–Physical–Social Perspective on Future Smart Distribution Systems," *Proc. IEEE*, pp. 1–31, 2022, doi: 10.1109/JPROC.2022.3192535.

[7] P. Voigt and A. von dem Bussche, *The EU General Data Protection Regulation (GDPR)*. Cham: Springer International Publishing, 2017. doi: 10.1007/978-3-319-57959-7.

[8] Standing Committee of the National People's Congress, "Data Security Law of the People's Republic of China." Jun. 10, 2021. [Online]. Available: https://gkml.samr.gov.cn/nsjg/bgt/202111/t20211105_336461.html

[9] A. Xu, Z. Zheng, F. Wu, and G. Chen, "Online Data Valuation and Pricing for Machine Learning Tasks in Mobile Health," in *IEEE INFOCOM 2022 - IEEE Conference on Computer Communications*, May 2022, pp. 850–859. doi: 10.1109/INFOCOM48880.2022.9796669.

[10] M. Yu *et al.*, "Pricing Information in Smart Grids: A Quality-Based Data Valuation Paradigm," *IEEE Trans. Smart Grid*, vol. 13, no. 5, pp. 3735–3747, Sep. 2022, doi: 10.1109/TSG.2022.3171665.

[11] C. E. Shannon, "A mathematical theory of communication," *Bell Syst. Tech. J.*, vol. 27, no. 3, pp. 379–423, Jul. 1948.

[12] B. Wang, Q. Guo, T. Yang, L. Xu, and H. Sun, "Data valuation for decision-making with uncertainty in energy transactions: A case of the two-settlement market system," *Appl. Energy*, vol. 288, p. 116643, Apr. 2021, doi: 10.1016/j.apenergy.2021.116643.

[13] A. Agarwal, M. Dahleh, and T. Sarkar, "A Marketplace for Data: An Algorithmic Solution," in *Proceedings of the 2019 ACM Conference on Economics and Computation*, in EC '19. New York, NY, USA: Association for Computing Machinery, Jun. 2019, pp. 701–726. doi: 10.1145/3328526.3329589.

[14] C. Gonçalves, P. Pinson, and R. J. Bessa, "Towards Data Markets in Renewable Energy Forecasting," *IEEE Trans. Sustain. Energy*, vol. 12, no. 1, pp. 533–542, Jan. 2021, doi: 10.1109/TSTE.2020.3009615.

[15] R. B. Myerson, "Optimal Auction Design," *Math. Oper. Res.*, vol. 6, no. 1, pp. 58–73, 1981.

[16] L. Han, J. Kazempour, and P. Pinson, "Monetizing Customer Load Data for an Energy Retailer: A Cooperative Game Approach," in *2021 IEEE Madrid PowerTech*, Jun. 2021, pp. 1–6. doi: 10.1109/PowerTech46648.2021.9494834.

[17] B. Wang, Q. Guo, and Y. Yu, "Mechanism design for data sharing: An electricity retail perspective," *Appl. Energy*, vol. 314, p. 118871, May 2022, doi: 10.1016/j.apenergy.2022.118871.

[18] L. D. Nguyen, I. Leyva-Mayorga, A. N. Lewis, and P. Popovski, "Modeling and Analysis of Data Trading on Blockchain-Based Market in IoT Networks," *IEEE Internet Things J.*, vol. 8, no. 8, pp. 6487–6497, Apr. 2021, doi: 10.1109/JIOT.2021.3051923.

[19] Q. Li *et al.*, "Database Watermarking Algorithm Based on Decision Tree Shift Correction," *IEEE Internet Things J.*, vol. 9, no. 23, pp. 24373–24187, Dec. 2022, doi: 10.1109/JIOT.2022.3188631.

[20] A. Botterud *et al.*, "Wind Power Trading Under Uncertainty in LMP Markets," *IEEE Trans. Power Syst.*, vol. 27, no. 2, pp. 894–903, May 2012, doi: 10.1109/TPWRS.2011.2170442.

[21] B. Wang, T. Xia, Q. Guo, L. Xu, and H. Sun, "Data Valuation in Electricity Transactions with Uncertainty Considering Risk Preferences," in *2022 IEEE 5th International Electrical and Energy Conference (CIEEC)*, May 2022, pp. 917–921. doi: 10.1109/CIEEC54735.2022.9846601.

[22] H. Zhao *et al.*, "Data-Driven Risk Preference Analysis in Day-Ahead Electricity Market," *IEEE Trans. Smart Grid*, vol. 12, no. 3, pp. 2508–2517, May 2021, doi: 10.1109/TSG.2020.3036525.

[23] H. Guo, Q. Chen, Q. Xia, and C. Kang, "Deep Inverse Reinforcement Learning for Objective Function Identification in Bidding Models," *IEEE Trans. Power Syst.*, vol. 36, no. 6, pp. 5684–5696, Nov. 2021, doi: 10.1109/TPWRS.2021.3076296.

[24] L. He, Y. Liu, and J. Zhang, "An Occupancy-Informed Customized Price Design for Consumers: A Stackelberg Game Approach," *IEEE Trans. Smart Grid*, vol. 13, no. 3, pp. 1988–1999, May 2022, doi: 10.1109/TSG.2022.3141934.

[25] Y. Li, B. Wang, Z. Yang, J. Li, and C. Chen, "Hierarchical stochastic scheduling of multi-community integrated energy systems in uncertain environments via Stackelberg game," *Appl. Energy*, vol. 308, p. 118392, Feb. 2022, doi: 10.1016/j.apenergy.2021.118392.

[26] A. Radovanovic, T. Nesti, and B. Chen, "A Holistic Approach to Forecasting Wholesale Energy Market Prices," *IEEE Trans. Power Syst.*, vol. 34, no. 6, pp. 4317–4328, Nov. 2019, doi: 10.1109/TPWRS.2019.2921611.

[27] D. H. Vu, K. M. Muttaqi, A. P. Agalgaonkar, and A. Bouzerdoum, "Short-Term Forecasting of Electricity Spot Prices Containing Random Spikes Using a Time-Varying Autoregressive Model Combined With Kernel Regression," *IEEE Trans. Ind. Inform.*, vol. 15, no. 9, pp. 5378–5388, Sep. 2019, doi: 10.1109/TII.2019.2911700.

[28] M. Afrasiabi, M. Mohammadi, M. Rastegar, L. Stankovic, S. Afrasiabi, and M. Khazaei, "Deep-Based Conditional Probability Density Function Forecasting of Residential Loads," *IEEE Trans. Smart Grid*, vol. 11, no. 4, pp. 3646–3657, Jul. 2020, doi: 10.1109/TSG.2020.2972513.

[29] W. Kong, Z. Y. Dong, D. J. Hill, F. Luo, and Y. Xu, "Short-Term Residential Load Forecasting Based on Resident Behaviour Learning," *IEEE Trans. Power Syst.*, vol. 33, no. 1, pp. 1087–1088, Jan. 2018, doi: 10.1109/TPWRS.2017.2688178.

[30] H. Oh, S. Park, G. M. Lee, J. K. Choi, and S. Noh, "Competitive Data Trading Model With Privacy Valuation for Multiple Stakeholders in IoT Data Markets," *IEEE Internet Things J.*, vol. 7, no. 4, pp. 3623–3639, Apr. 2020, doi: 10.1109/JIOT.2020.2973662.

[31] M. Hodge, "Optimization of conditional value-at-risk," *J. Risk*, Mar. 2000, Accessed: Nov. 27, 2022. [Online]. Available: https://www.risk.net/node/11441

[32] C. J. Dent, J. W. Bialek, and B. F. Hobbs, "Opportunity Cost Bidding by Wind Generators in Forward Markets: Analytical Results," *IEEE Trans. Power Syst.*, vol. 26, no. 3, pp. 1600–1608, Aug. 2011, doi: 10.1109/TPWRS.2010.2100412.

[33] E. Baeyens, E. Y. Bitar, P. P. Khargonekar, and K. Poolla, "Coalitional Aggregation of Wind Power," *IEEE Trans. Power Syst.*, vol. 28, no. 4, pp. 3774–3784, Nov. 2013, doi: 10.1109/TPWRS.2013.2262502.

[34] J. Wang *et al.*, "Incentive mechanism for sharing distributed energy resources," *J. Mod. Power Syst. Clean Energy*, vol. 7, no. 4, pp. 837–850, Jul. 2019, doi: 10.1007/s40565-019-0518-5.